\documentclass[referee, pdflatex,sn-mathphys-num]{sn-jnl}

\usepackage{graphicx}%
\usepackage{multirow}%
\usepackage{amsmath,amssymb,amsfonts}%
\usepackage{amsthm}%
\usepackage{mathrsfs}%
\usepackage[title]{appendix}%
\usepackage{xcolor}%
\usepackage{textcomp}%
\usepackage{manyfoot}%
\usepackage{booktabs}%
\usepackage{algorithm}%
\usepackage{algorithmicx}%
\usepackage{algpseudocode}%
\usepackage{listings}%

\theoremstyle{thmstyleone}%
\theoremstyle{thmstyletwo}%
\def\F {\boldsymbol{\Phi}}
\def\x {\boldsymbol{x}}
\newcommand{\ie}{{\it i.e.},~}

\theoremstyle{thmstylethree}%

\begin{document}

\title[Article Title]{4D \textit{operando} X-ray nano-holo-tomography reveals multiscale chemomechanics in Silicon-Graphite anode}

\author*[1]{\fnm{Victor} \sur{Vanpeene}}\email{victor.vanpeene@cea.fr}

\author[2]{\fnm{Olga} \sur{Stamati}}\email{olga.stamati@3sr-grenoble.fr}

\author[3]{\fnm{Francois} \sur{Cadiou}}\email{francois.cadiou@cea.fr}

\author[1]{\fnm{Quentin} \sur{Jacquet}}\email{quentin.jacquet@cea.fr}

\author[4]{\fnm{Julie} \sur{Villanova}}\email{julie.villanova@esrf.fr}

\author*[1]{\fnm{Sandrine} \sur{Lyonnard}}\email{sandrine.lyonnard@cea.fr}

\affil*[1]{\orgname{Univ. Grenoble Alpes, CEA, CNRS, Grenoble INP, IRIG, SyMMES}, \orgaddress{ \city{Genoble}, \postcode{38000}, \country{France}}}

\affil[2]{\orgname{Univ. Grenoble Alpes, CNRS, Grenoble INP, 3SR}, \orgaddress{\city{Genoble}, \postcode{38000}, \country{France}}}

\affil[3]{\orgname{Univ. Grenoble Alpes, CEA, Liten, DEHT}, \orgaddress{\city{Genoble}, \postcode{38000}, \country{France}}}

\affil[4]{\orgname{ESRF - The European Synchrotron}, \orgaddress{\city{Genoble}, \postcode{38000}, \country{France}}}


\abstract{Linking electrode microstructure to electrochemical performance is essential for optimizing Li-ion batteries. However, this requires mechanistic 4D observations at ultimate spatio-temporal scales, which remains elusive. Here we demonstrate the use of \textit{operando} synchrotron X-ray nano-holo-tomography combined with Digital Volume Correlation to track chemomechanical dynamics at both particle (local) and electrode (averaged) scales. Quantitative scale-bridging image analysis is applied to a high-capacity silicon-graphite anode during its formation cycle. Our findings reveal that local diffusion properties, graphite particle morphology and position in the electrode, distance to silicon clusters, surface contact with electrolyte and mechanical deformations, all have a direct impact on the local electrochemical activity and irreversibility - but these parameters are not equally important. Particularly, we identify fast diffusion channels that play a key role and counterbalance intrinsic depth-dependent reaction heterogeneities due to ionic/electronic diffusion limitations. The various structural factors that determine Gr-Si battery performance beyond ensemble properties are classified using a scale of influence, providing a practical framework for the optimization of materials and electrode manufacturing.}




\maketitle

\section{Introduction}\label{sec1}
Degradation and irreversible morphological changes in Li-ion battery electrodes are complex intertwinned phenomena, involving cracking and disconnections of particles, pore clogging, solid electrolyte interphase (SEI) growth, localized strains and deformations, resulting overtime in continuous loss of cyclable lithium and anticipated cell failure.\cite{dominko2021} The heterogeneity and amplitude of these degradation mechanisms mainly depend on the electrode microstructure and active material properties, highlighting the need to reversely optimize the electrode microstructures to achieve the required electrochemical performance and durability. However, precisely controlling and tailoring the porous microstructure of an electrode is very difficult. It depends on both the mixed materials - active particles, binders, coatings and additives - and the manufacturing process,\cite{CHOUCHANE2019227285} from slurry preparation, film casting, drying, calendaring, to cell assembly.\cite{Franco2019, Franco2021} Consequently, current state-of-the-art electrode optimization relies on tuning empirically a limited set of averaged parameters, such as active particle size distribution, volume fractions, macroscopic porosity, averaged tortuosity and electrode thickness. These parameters have a direct influence on electronic and ionic transport/transfer mechanisms, thereby affecting the efficiency and reversibility of electrochemical reactions. However, they do not capture the 3D multiscale complexity of electrode structures, their evolving topology and local mechanics, as well as the spatial distribution of solid and liquid phases. Therefore, to establish the structure-performance relationship beyond trial-and-error approaches, understanding the intricate chemomechanical interactions and how they regulate electrochemical activity is crucial.\cite{deVasconcelos2022}

Analysing the coupling between structure, morphology, stress,  reactions and transport pathways requires advanced methodologies to simultaneously monitor the electrode dynamics and chemomechanical evolution at both particle (local) and electrode (averaged) scales. 
Experimentally, this is challenging due to (i) the need to spatially resolve heterogeneities spanning over multiple length scales, (ii) the difficulty to detect and quantify poorly-contrasting phases, typically porous network, binder and other organic species/layers,\cite{Cadiou2020-li} and (iii) the evolving nature of the microstructure during cycling. Hence, 3D space-resolved approaches with ultimate/tuneable resolutions are needed to determine the electrode hierarchical structure,\cite{Scharf2022, Lu2020} in combination with representative \textit{operando} conditions to capture the dynamic changes during electrode (de)lithiation.\cite{Liu2019}

Several techniques are available, each typically specializing in tackling one of the three aforementioned challenges - but none of them is capable of taking them simultaneously.
High resolution studies probing heterogeneities at the particle scale were reported by STEM-EELS,\cite{Boniface2016, JiahanLestriez2022} optical microscopy,\cite{Pandya2023_nat, Pandya2023_nanolett,OptMicro_raman_HWLee2023} STXM,\cite{Li2020, Zhao_Bazant2023} Ptychography,\cite{Wood_multimodal_nano} XRD-CT\cite{Finegan2019, Pietsch2016_xrd_ct} or BCDI\cite{Martens2023, Vostrov2024}. They provided key insights into lithiation mechanisms by imaging reaction front propagations, interfacial transfer mechanisms, local strains and phase/species spatial distributions. However, with these techniques, only limited selections of active particles in given states of charge (SoC) are randomly measured, raising concerns about representativeness and universality of the findings. FIB-SEM has also been used to probe electrode microstructures in 3D with nanoscale resolution, analyzing SEI and particle morphology evolutions induced by charging and discharging\cite{Liu2025, Perrenot2024, SiSiC_slag_batteries2025}. Coupling with chemical information was brought by using complementary techniques as ToF-SIMS\cite{Bordes2016, Korsunky2015} and XPS.\cite{Marcus2010} Nevertheless, the probed volumes remain small due to hours-long acquisition time and the technique itself is quasi exclusively restricted to \textit{ex situ} analyses, as FIB milling is destructive.

Alternatively, X-ray computed tomography (CT) techniques offer the advantage of being non-invasive, 3D by nature, versatile towards analyzing a wide range of chemistries,\cite{Scharf2022, pietsch2017, Vanpeene2025} probing representative volumes and reaching excellent spatio-temporal resolutions when used at a synchrotron facility (from \textit{$\mu$m} down to several hundreds of \textit{nm}, from hours to minutes). Synchrotron micro-CT has been widely used in the battery community primarily at the cell and component level.\cite{pietsch2017, bond_dahn2025, Finegan2015} Recent studies reported defects in cylindrical geometries due to manufacturing or aging,\cite{Erik_cylindrical, Shearing_cylindrical_2024} geometry-dependent electrolyte motions,\cite{bond_dahn2025} component distortions and delaminations,\cite{Shearing_cylindrical_2024} \textit{etc}. In combination with far-field high-energy diffraction microscopy, micro-CT was also used to monitor grain-level strain responses in garnet-type solid electrolytes, revealing a stochastic failure mechanism affected by local microstructural heterogeneity. \cite{Dixit2022} Advanced chemomechanical quantification can be also obtained by applying 3D image analysis techniques in timeseries CT data. For example, Digital Volume Correlation (DVC),\cite{bay1999digital, GEERS19964293} allows for the measurement of 3D kinematic fields, displacements and the derived strains, which can in turn be decomposed into an isotropic (\ie spherical) and deviatoric part, enabling analysis of volume and shear distortions. The DVC approach has been applied to different types of battery electrode materials, such as ${LiMn_{2}O_{4}}$\cite{Eastwood2014} and graphite\cite{Pietsch2016_tomo_dvc} yielding information about mechanical failure modes at mesoscopic scales. More recently Valisammagari \textit{et al.}\cite{abi_dvc_2024} applied it to Gr-Si electrodes at the mesoscopic scale and found a relationship between the local strain distribution and the electrode microstructure, but the restricted spatial resolution used did not allow establishing a comprehensive relationship between local microstructure rearrangement and electrode activity/reversibility.  

Despite these advances, reported works at the electrode/cell level generally lack submicron information due to their limited spatial resolution, which is mandatory for particle-scale analysis. X-ray nano-computed tomography\cite{Scharf2022} overcomes this limitation. In particular, the propagation-based phase-contrast approach\cite{holo_cloetens_1999, Zabler2005} (so-called X-ray nano-holo-tomography, nano-CT) provides phase segmentation capabilities for low-attenuating materials, relevant for electrodes made of light elements such as graphite/silicon composites,\cite{Vanpeene2025, Wood_multimodal_nano, TuTu2020} for which absorption-based tomography fails to resolve with sufficient contrast the difference between the material phases. Moreover, nano-CT provides a unique combination of a large field of view (typically several tens of microns) and high spatial resolution ($\leq$ 200 nm),\cite{Langer2012} enabling to resolve features from the entire electrode scale down to individual particles.\cite{id16a_monaco, Martens_Vanpeene2023} Different techniques of X-ray nano-computed tomography have been extensively employed for post-mortem characterization of electrodes revealing changes in porous network size, connectivity and tortuosity\cite{CHENWIEGART2014349, vanpeene2025_adv, Frisco2016-rb, ABREGOMARTINEZ2023118004} and at the particle/electrode scale\cite{ZHAO2018, CHENWIEGART201258}, or used to provide real microstructures for image-based simulations.\cite{Lu2020, ZChen2017_LMnO, Yan2012-lk} Only recently, fast acquisition times (below tens of minutes) at the sub-micron became achievable at $4^{th}$ generation synchrotrons thanks to the extreme source brilliance, enabling a ca. 10 fold gain in spatiotemporal resolution. This breakthrough has opened the way for 4D (3D + time) nano-CT acquisitions to study \textit{in-situ}/\textit{operando} structural changes during cycling with representative sample size/preparation, provided that custom battery cells and compatible experimental set-ups are developed \cite{Vanpeene2025}, as well as complex experimental workflows designed to preserve sample integrity, focus on the areas of interest, combine large (electrode-scale) field of view with extreme (nanoscale) resolution, and ensure representativeness and reproducibility at each measurement step.  

Here, we tackle this combination of challenges and demonstrate a methodology for multiscale \textit{operando} nano-CT measurements combined with DVC, applied to a high-performance graphite-silicon composite electrode during its formation cycle. This multiphase composite material is an ideal candidate for probing chemomechanical coupling at electrode and particle scales, since it is well known that Gr-Si anodes undergo heterogeneous volume changes during (de)lithiation due to the alloying process of silicon with lithium (300\%wt expansion in bulk).\cite{Obrovac2014, Choi2016}, and, in parallel, the (de)lithiation mechanism is sequential, with silicon being lithiated first and delithiated after graphite.\cite{Berhaut2023, OptMicro_raman_HWLee2023} By analyzing individual particle behaviors as well as subgroups of particles, together with electrode-averaged microstructural features, we quantify and track local and ensemble micromechanical states. We show that the electrode's deformation is governed on average by the graphite particles, while its irreversibility arises from the local heterogeneity in the silicon material distribution throughout the electrode, leading to local rearrangements. We classify the several competing structural factors that regulate the Gr-Si battery performance introducing a scale of impact that helps to showcase future pathways towards electrode optimization.

\section{Results}\label{sec2}
\subsection{4D multiscale X-ray nano-imaging}\label{subsec2}
\begin{figure}[h!]
\centering
  \includegraphics[width=10cm]{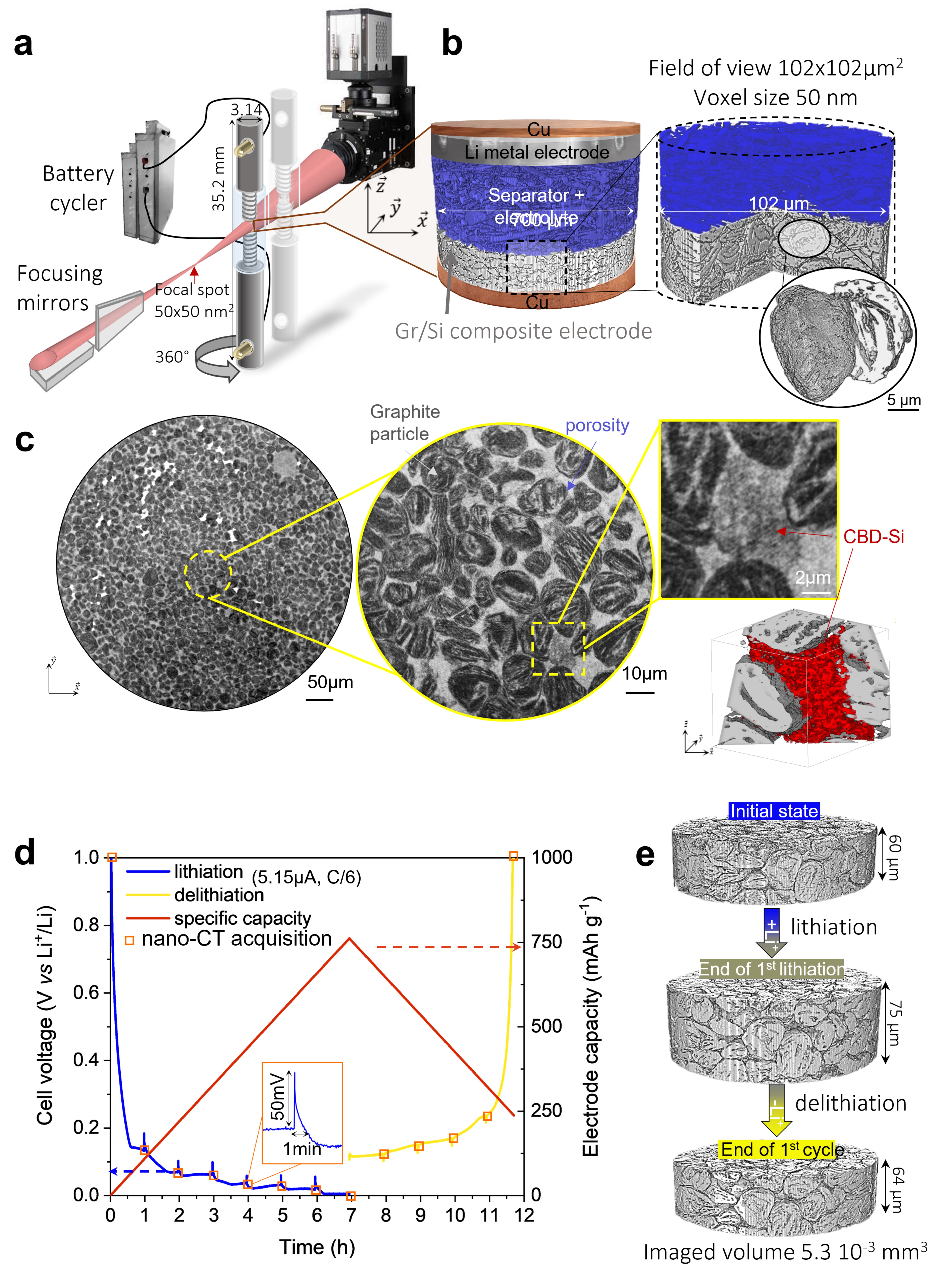}
  \caption{\textbf{\textit{Operando} X-ray nano-holo-tomography (nano-CT) of Gr-Si composite electrode.} a) Sketch of the \textit{operando} nano-CT setup at the ID16B beamline of ESRF featuring the electrochemical cell housed in a dedicated holder and connected to a potensiostat. b) 3D rendered images of the analyzed half-cell assembly (Li/separator/Gr-Si composite electrode) imaged at 257 nm voxel size and zoom-in rendering at high magnification (50 nm voxel size) focused on the electrode and particle scale. c) Nano-CT reconstructed grayscale slices acquired in \textit{operando} conditions at different magnifications (257 and 50 nm voxel size) with a close-up view of one large CBD-Si cluster in 2D and 3D. d) First formation cycle of the composite electrode (5.15 ${\mu}$A, C/6) with voltage and capacity evolution over time, along with the hourly acquisition steps indicated by the orange rectangles and e) 3D renderings of the reconstructed images of the Gr-Si electrode (50 nm voxel size) at specific time steps (fresh state, end of $1^{st}$ lithiation, end of $1^{st}$ cycle).}

  \label{setup}
\end{figure}

X-ray nano-imaging was performed on a calendared industry-grade manufactured composite electrode consisting of two active materials, graphite and nano-silicon (74 and 11\%wt respectively), mixed with binder and additives. Details on the materials and electrode fabrication are given in Methods. Sample preparation, cell configuration and \textit{operando} imaging setup were optimized and standardized in order to reach the small sample size required for both high resolution and high contrast imaging, while allowing a reliable electrochemical and 3D nano-imaging monitoring in half-cell configuration.\cite{Vanpeene2025} The dedicated setup comprised a custom 1.5 mm diameter miniature cell held under constant pressure with the Gr-Si electrode as working electrode separated from the lithium foil reservoir by a separator soaked in the electrolyte (Fig. \ref{setup}a, see Methods for details on the electrochemical testing, cell assembly and synchrotron measurement protocols). 

First, multiscale X-ray nano-CT conditions were tuned to select the best representative volume, \ie best compromise  between spatial/temporal resolution and field of view under \textit{operando} conditions (Fig. \ref{setup}b, Supplementary Fig. S1). Typically, a large area is first imaged at low resolution before zooming in at an increased resolution. We determined that a probed volume of $102\times102\times102 {{\mu}m}^{3}$ (voxel size of 50 nm) was representative of the overall electrode microstructure (see Methods and Supplementary Note 1, Fig. S2). The examined microstructure of the composite electrode displays a multiscale and heterogeneous morphology attributed to the distinct characteristics of the active particles. Graphite particles are large (average 12 ${\mu}$m diameter) and porous (Fig. \ref{setup}b), while the silicon nanoparticles (nominal size 200 nm) tend to aggregate, forming several large regions of carbon-binder-silicon (CBD-Si) domains, evidenced in the \textit{in situ} nano-CT reconstructed volumes at both low and high magnifications (Fig. \ref{setup}c and Supplementary Fig. S2a-b). These large clusters of CBD-Si-rich phase are randomly distributed throughout the electrode and range from 5 to 15 ${\mu}$m in size. A key asset for our study is the quality of nano-CT data acquired in \textit{operando} conditions where pore filling by electrolyte and cell housing can produce artefacts or image blurring. Nevertheless, the reconstructed image quality in \textit{operando} conditions was globally preserved with respect to the \textit{ex situ} measurement, achieving good contrast in areas where large clusters of CBD-Si are identified, thanks to the phase contrast capabilities of holo-tomography, along with the high X-ray transmission and the very stable customized setup used for imaging (Fig. S3).

The dispersion of silicon particles in the composite microstructure is highly inhomogeneous at the electrode scale, as revealed by complementary 2D mappings performed by \textit{operando} Wide Angle X-ray Scattering (WAXS) using a $5\times20 {\mu}m^{2}$ beam and 5(V)$\times$20(H) ${\mu}m^{2}$ scanning resolution (see Methods and Supplementary Note 2, Fig. S4). Silicon is widespread within the pristine electrode (\textit{e.g.}, the silicon diffraction peak (111) is measured throughout the entire microstructure), but certain high concentration regions are localized, which likely correspond to the large CBD clusters observed in the nano-CT reconstructed volumes (Supplementary Fig. S1c, d). The measured volume fraction from the large CBD-Si clusters is 4.1 $\%^{v}$, against a theoretical volume fraction of 18 $\%^{v}$ expected from the anode composition. Therefore, a substantial portion (75\%) of the silicon material is intimately mixed with graphite particles and/or grafted on their surface, remaining unresolved within the nano-CT volume, due to the limited spatial resolution ($\leq$ 150 nm) compared to the average silicon particle size (200 nm). Consequently, quantitative silicon analysis is not feasible at individual particle/aggregate scale. Nonetheless, the silicon (de)lithiation mechanism can be indirectly monitored by investigating its impact on the graphite, through a detailed particle-by-particle morphological analysis, as will be shown in next sections. Using a machine learning segmentation of the reconstructed grayscale images (see Supplementary Note 3), both the graphite particles and the electrode porosity can be isolated across the electrode (Fig. \ref{electrode_scale}a), enabling a deeper investigation of the multiphase reaction mechanisms, electrode deformation, and pore network dynamics.

Once the tailored sample preparation, cell operation, and X-ray data acquisition protocols were validated, interrupted \textit{operando} X-ray nano-tomography was performed under these representative conditions, and electrode-scale dynamics were successfully measured during cycling within minutes of temporal resolution. Holo-tomography scans consisted of 4 distances acquisition (total acquisition time of 9 min), acquired every hour during the Gr-Si composite electrode formation cycle performed with constant current (C/6) in the nano-CT custom half-cell with no lithium limitation (Fig. \ref{setup}d). Open circuit potential steps, \ie where no current are applied on the system, were used to monitor the half-cell voltage evolution during the X-ray exposure times and control beam damage. Attenuated and reversible voltage perturbations were monitored during X-ray exposure steps, along with a reliable coulombic efficiency for the electrode formation cycle, matching the value measured for a reference cell cycled outside the beam (78 vs. 80 \%, Supplementary Note 4, Fig. S5). Moreover, the radiation doses received by the electrolyte and active material during the full experiment were respectively 0.52 and 0.39 MGy, falling below the previously identified dose threshold of 1.5 MGy,\cite{Vanpeene2025} hence insuring that beam interactions did not alter the reaction process kinetics nor damaged the sample.  By examining the timeseries nano-CT images, changes in the electrode thickness can be visually demonstrated, \textit{e.g.} swelling during lithiation followed by contraction during delithiation (60 ${\mu}$m to 75 ${\mu}$m, and back to 64 ${\mu}$m, Fig. \ref{setup}e), indicating the expected material’s behavior during the (dis)charging process. This confirms the reliability of the setup, and hence enables further quantitative investigation of the underlying dynamic mechanisms through complementary morphological, mechanical, and electrochemical analyses within the electrode down to the particle scale. 

\subsection{Morphological evolution of the microstructure}\label{subsec3}
\begin{figure*}[h!]
\centering
  \includegraphics[width=10cm]{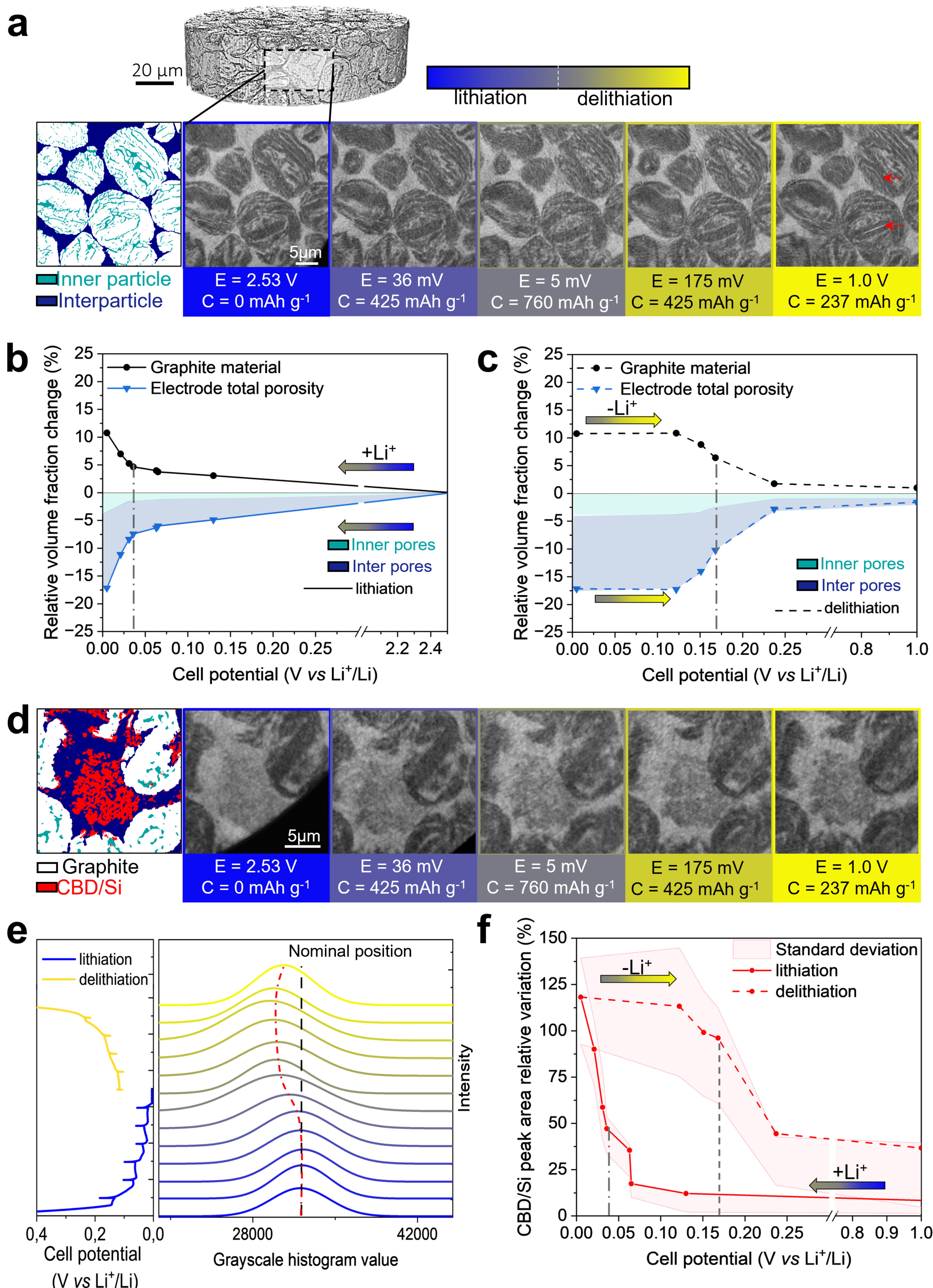}
  \caption{\textbf{Gr-Si electrode microstructural evolution along the formation cycle.} Quantification of graphite and porosity evolution on a ${102 \times 102 \times 102}{{\mu}m}^{3}$ volume within the electrode: a) Illustration of the 3D volume analyzed for the quantitative analysis of b,c, and close up view of segmentation with reconstructed in-plane nano-CT slices at specific cycling steps. The red arrows indicate the formation of new pores inside the graphite particles. Evolution of the relative volume fraction change over the cell potential for the graphite particles in black and global porosity (dark blue and cyan corresponds to interparticle and inner graphite particle porosity respectively) during b) lithiation and c) delithiation. Focus on the silicon material: d) Close up views of segmentation and reconstructed in-plane nano-CT slices of a selected CBD-Si cluster at specific cycling steps and e) evolution of the 3D cluster grayvalue histograms during cycling; f) relative variation versus cell potential of the CBD-Si rich domain (labeled in red in d) peak integrated from the grayvalue histogram.}
  
  \label{electrode_scale}
\end{figure*}

Relative changes in both the graphite and porosity volume fractions indicate a multi-step sequential (de)lithiation process (Fig. \ref{electrode_scale}a and Supporting Video S1). Initially, during lithiation, the graphite material fraction and porosity stay balanced over a large potential range ($\geq$ 60 mV \textit{vs.} $Li^{+}/Li$), leading to an overall swelling of the electrode microstructure. Then, below 35 mV \textit{vs.} $Li^{+}/Li$, a sudden decrease in porosity is measured, extending from -7.5 to -17\%  (see dash line in electrode total porosity curve, Fig. \ref{electrode_scale}b). During delithiation, the porosity progressively increases, with a larger step from -10 to -2.5\% after reaching 175 mV \textit{vs.} $Li^{+}/Li$ (see dash line in electrode total porosity curve, Fig. \ref{electrode_scale}c). This observation is consistent with a sequential process, with graphite delithiation occurring first, followed by that of silicon, as already reported in literature.\cite{Berhaut2023}
The global porosity network can be divided into two main components: interparticle porosity, associated with the electrode microstructure; and intraparticle porosity (\textit{e.g.}, inner-particle pores) characteristic of the active material (Fig. \ref{electrode_scale}a). Initially we find a global macro-porosity of $38\%^v$, composed of $13\%^v$ inner-particle and $25\%^v$ interparticle porosity. This global porosity value is in agreement with the material specifications after electrode calendaring (33\%). Distinguishing inter- and intra-porosities provides valuable information about the nature and function of the pore network beyond a simple global behavior, enabled by the high spatial resolution of the nano-CT. The contributions coming from the inter and inner-particle porosity are shown as dark and light blue colored areas in the electrode total porosity curves of Fig. \ref{electrode_scale}b,c. During the initial swelling phase ($\geq$ 50-60 mV \textit{vs.} $Li^{+}/Li$), both the interparticle and inner-particle porosity follow the trend of the macro-porosity (continuous line), with 40\% of the contribution coming from the inner pores. After the abrupt macro-porosity decrease noticed once the system reaches the silicon activity domain, the inter- and inner-particle porosities drop down to respectively -15 and -22\% relative volume fractions (Fig. S6a). At that stage, the contribution coming from the inner pores is lower (30\%, Fig. \ref{electrode_scale}c). However, during delithiation, the inner particle porosity progressively increases, until eventually exceeding the levels in the fresh state (+18 \%, Fig. S6a), indicating the expansion of the inter-sheet spaces and the formation of new pores in the active material, which can be observed in Fig. \ref{electrode_scale}a (red arrows), \ie graphite sheets exfoliation. In fact, the average local pore size in graphite particle is 420 nm at the end of the formation cycle versus 280 nm initially (Fig. S6b). The contribution coming from the inner pores is now up to 50\% (Fig. \ref{electrode_scale}c), which is essential to consider. This important contribution is usually not seen by micro-CT techniques and/or under-looked in general due to lack of spatial resolution. These freshly opened surfaces in contact with the electrolyte are prone to additional SEI formation in the subsequent cycle, resulting in cumulative Li inventory loss. Additionally, a significant reduction of the interparticle porosity is detected (-15 \%), which could be associated with incomplete graphite delithiation and/or pore clogging induced by silicon activity (Fig. S6a,b).\cite{Berhaut2023, Von_Kessel2024-kg, Moon2021-xo}

Meanwhile, progressive changes in grayscale values were observed within the large clusters of CBD-Si, as illustrated by the reconstructed slices at different time steps in Fig. \ref{electrode_scale}d. As already mentioned, the behavior of these clusters does not represent all silicon particles, but rather localized regions of highly concentrated aggregates. Within the examined cluster area, evidence of irreversibility appears as the grayscale value shifts toward lower (darker) intensities (Fig. \ref{electrode_scale}e), compared to the reference level (chosen as the one measured in electrolyte-filled regions in the separator, see Supplementary Note 3 and Fig. S7). This relative shift could be the result of enhanced SEI formation in these domains and/or incomplete delithiation of silicon particles. To monitor these changes, the evolution of the integrated peak area extracted from the grayscale images of Fig. \ref{electrode_scale}d is followed during the formation cycle (Fig. \ref{electrode_scale}f), serving as a metric to indirectly track the change in Si volume after (de)lithiation. The results reveal a large volume increase in the CBD-Si region studied, up to 120\% at the end of lithiation, with significant irreversible swelling (+ 30\%) remaining at the end of the formation cycle. The evolution profile highlights distinct voltage steps for the (de)lithiation of silicon, further supporting our previous observations made at the electrode scale based on graphite particle analysis. In addition, the kinetics and transition of graphite lithiation staging process relative to silicon activity were monitored through WAXS mapping acquisitions during lithiation (Supplementary Fig. S4). It was observed that Si primarily reacts when the Stage2/2L is pure (capacity 600 mAh $g^{-1}$), and to a lesser extent when the graphite is fully lithiated (capacity 800 mAh $g^{-1}$), in agreement with both our nano-CT data and previous scattering-based works.\cite{Berhaut2019-ht} 

In conclusion, the global Gr-Si electrode microstructure exhibits general inflation during the first part of lithiation before suddenly densifying towards the end of lithiation, and then it contracts sequentially during the delithiation step. Capturing this sequence at the nanoscale further confirms the high reliability of our data acquisition and analysis workflow. The contraction is associated with, first, the increase of the inner graphite particle porosity through the opening of new pores, while the interparticle porosity continues decreasing, when both active phases are uptaking lithium, showing possible microstructural rearrangements. Subsequently, the interparticle porosity increases back when the silicon material delithiates, along with localized irreversible changes mostly concentrated in the large CBD-Si clusters. 

\subsection{Chemomechanical evolution at the electrode scale}\label{subsec4}
\begin{figure*}[h!]
\centering
  \includegraphics[width=10cm]{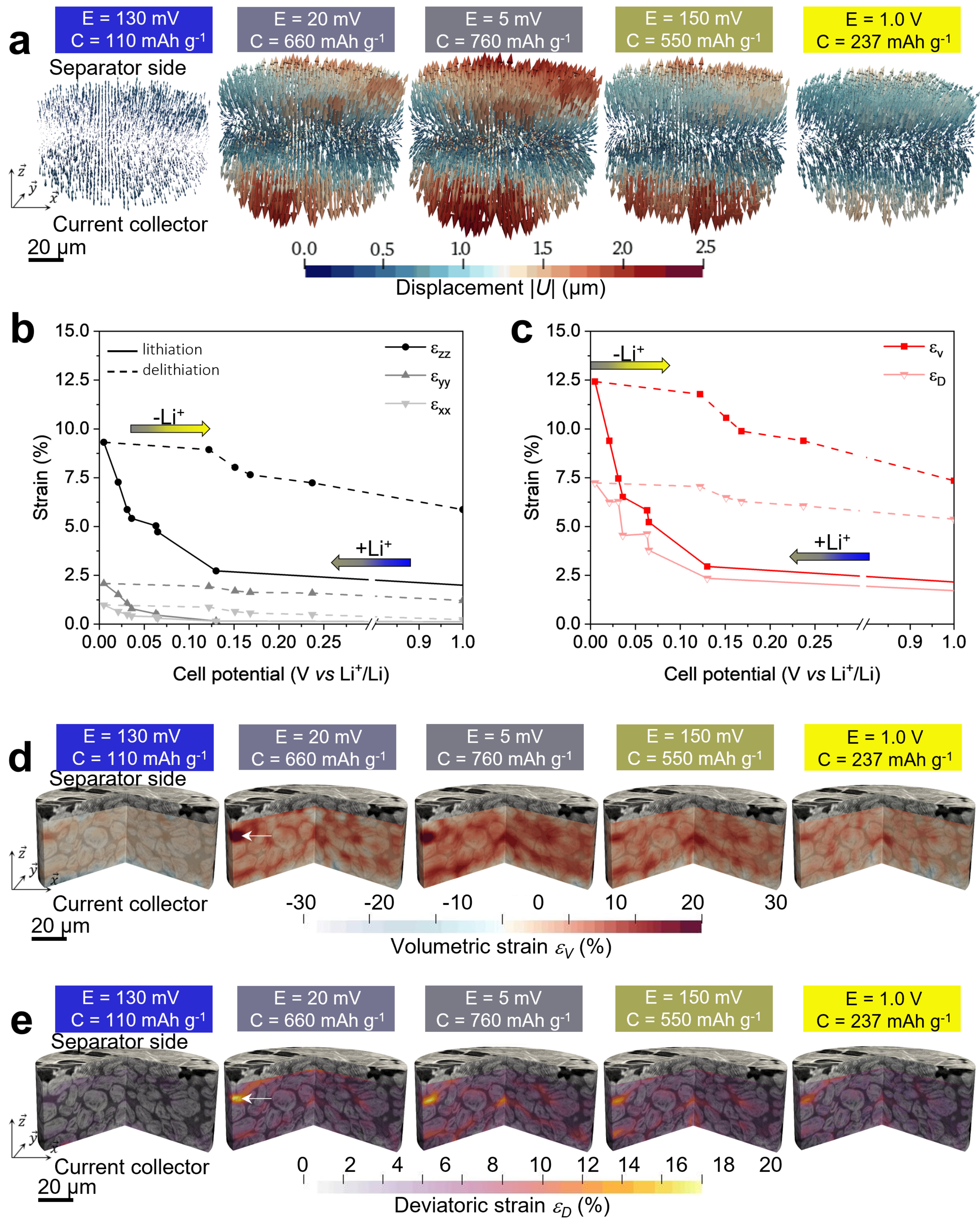}
  \caption{\textbf{Kinematic fields during the formation cycle obtained by local Digital Volume Correlation.} a) 3D renderings of the displacement fields of the electrode and evolution over the cell potential of the mean strain components b) along the three orthogonal directions and c) the two strain invariants volumetric ($\epsilon_{v}$) and deviatoric ($\epsilon_{D}$). 3D renderings of the d) total volumetric and e) deviatoric strain maps at specific cycling steps during lithiation (blue to brown) and delithiation (brown to yellow). Note that the rigid-body motion measured at the centre of mass of the electrode was subtracted and that for visualization the displacement vectors are scaled up 10 times.}
  \label{dvc_local}
\end{figure*}

To further understand the link between mechanics and chemical changes in the electrode, we need to resolve the spatial distribution of Gr, Si and pore network heterogeneities during cycling, and determine their impact on the mechanical behavior of the composite electrode. To this end, a detailed quantitative analysis of the electrode's microstructural deformation was performed by employing the 3D full-field measurement technique of Digital Volume Correlation, where sub-volumes of $4\times4\times4~\mu\mathrm{m}^{3}$ were locally tracked during the formation cycle, allowing the measurement of the 3D displacement fields, along with the derived strain fields in the electrode at the nanoscale (see Methods). The displacement maps clearly indicate that the electrode expands/contracts as a whole from its center of mass towards the separator and current collector sides during lithiation/delithiation (Fig. \ref{dvc_local}a and Supporting Videos S1,S2). The mean values of the measured orthogonal strain components demonstrate that the electrode's microstructural deformation is highly anisotropic, with an amplitude largely dominated by the component along its thickness (${\epsilon}_{zz}$ = 8.8\%), in agreement with previous reports.\cite{abi_dvc_2024, VANPEENE2019799} Regarding the in-plane deformation components, an asymmetric behavior in magnitude is observed (${\epsilon}_{xx}$ = 1.24\%, ${\epsilon}_{yy}$ = 2.43\%), highlighting rearrangements in the electrode microstructure large enough to disrupt the diffusion pathways (Fig. \ref{dvc_local}b). Notably, the axial deformation component ${\epsilon}_{yy}$ exhibits a large irreversibility, a trend also reflected by volumetric and deviatoric strains (Fig. \ref{dvc_local}c). At the onset of lithiation, both the graphite and porosity volume fractions increase at a similar rate (Fig. \ref{electrode_scale}b), leading to a progressive rise of volumetric strain (Fig. \ref{dvc_local}b), which intensifies as silicon lithiation starts to predominate below 35 mV \textit{vs.} $Li^{+}/Li$ (Fig. \ref{electrode_scale}f). At this stage, in-plane strains increase, showing the impact of silicon activity on the 3D microstructural rearrangement of the composite electrode. 

Furthermore, the deviatoric strain provides access to the shearing mechanism within the electrode volume. It remains lower in magnitude compared to the volumetric component, but evolves more progressively (Fig. \ref{dvc_local}c), presenting a staircase profile during lithiation with steps at 70, 35 and 20 mV \textit{vs.} $Li^{+}/Li$. These potentials correlate with the three main voltage plateaus observed in the cycling profile of Fig. \ref{setup}c, suggesting that they could correspond to the successive lithiation stages of graphite,\cite{Weng2023} shifted due to the higher electrode polarization associated with the silicon material. During these voltage plateaus, the lithiation level of individual graphite particles in the electrode is heterogeneous,\cite{Berhaut2023} leading to elevated anisotropic deformation resulting in localized shearing of the electrode microstructure. Moreover, a significant irreversibility (74\%) of the deviatoric contribution is found at the end of the formation cycle. These observations show that the formation step is critical not only for an efficient SEI formation, but also for an essential "preparation" of the electrode's morphology via in-plane and through-plane microstructural rearrangements.

In addition to the evolution of the mean values of the strain components, DVC allowed monitoring their spatial distribution at the electrode level (Fig. \ref{dvc_local}d,e and Supporting Videos S3,S4). Note that the correlation window size used falls well beyond the values reported in standard mesoscale DVC works windows in the range 20-150 ${\mu}m$ are usually employed,\cite{abi_dvc_2024, Pietsch2016_tomo_dvc, Eastwood2014} failing to resolve the grain-grain interfaces. In contrast, our method probes extremely localized discretized strain fields enabling views into grain interfaces and their role in the dynamical process at stake. At this scale, both the volumetric and deviatoric strain fields are shown to be rather homogeneous. Nevertheless, some hot spots for mechanical strain are identified throughout the electrode volume (seen as red regions in Fig. \ref{electrode_scale}d). Resolving and 3D localizing these areas is of high importance, as they can be regions of accelerated aging or degradation. In the present case, high localized volumetric strain is often associated with significant shearing, which is observed in particular in the area where a large CBD-Si cluster has previously been identified (white arrows in Fig. \ref{dvc_local}d,e and Fig. \ref{electrode_scale}d). Additionally, elevated shear strains are observed at graphite particles interfaces, where rearrangement is expected to be the highest. High shearing is also detected closer to the electrode/separator interface, attributed to mechanical incompatibility between the compliant separator and the dynamically expanding/contracting electrode.
This could further lead to particle disconnections at the electrode/separator interface and accelerated/enhanced aging of the top of the electrode, as also suggested in the literature.\cite{oney2025graphite} Although the imaging (and subsequent tracking) of the electrode/current collector interface was not possible here due to the limited field of view, we can anticipate that similar localized shear trends are exacerbated at the electrode/current collector interface, due to the rigid nature of copper, eventually leading to partial disconnection/delamination of the electrode, in agreement with earlier reports.\cite{abi_dvc_2024} Altogether, our nanoscale-resolved DVC approach enables to quantify mechanically-induced morphological changes with extreme local accuracy and reveal hot spots where local deformations are concentrated, which are expected to affect the electrochemical performance, particularly the efficiency of ionic diffusivity within the microstructure.

\subsection{Dynamics of ionic diffusion pathways}\label{subsec5}
\begin{figure*}[h!]
\centering
  \includegraphics[width=10cm]{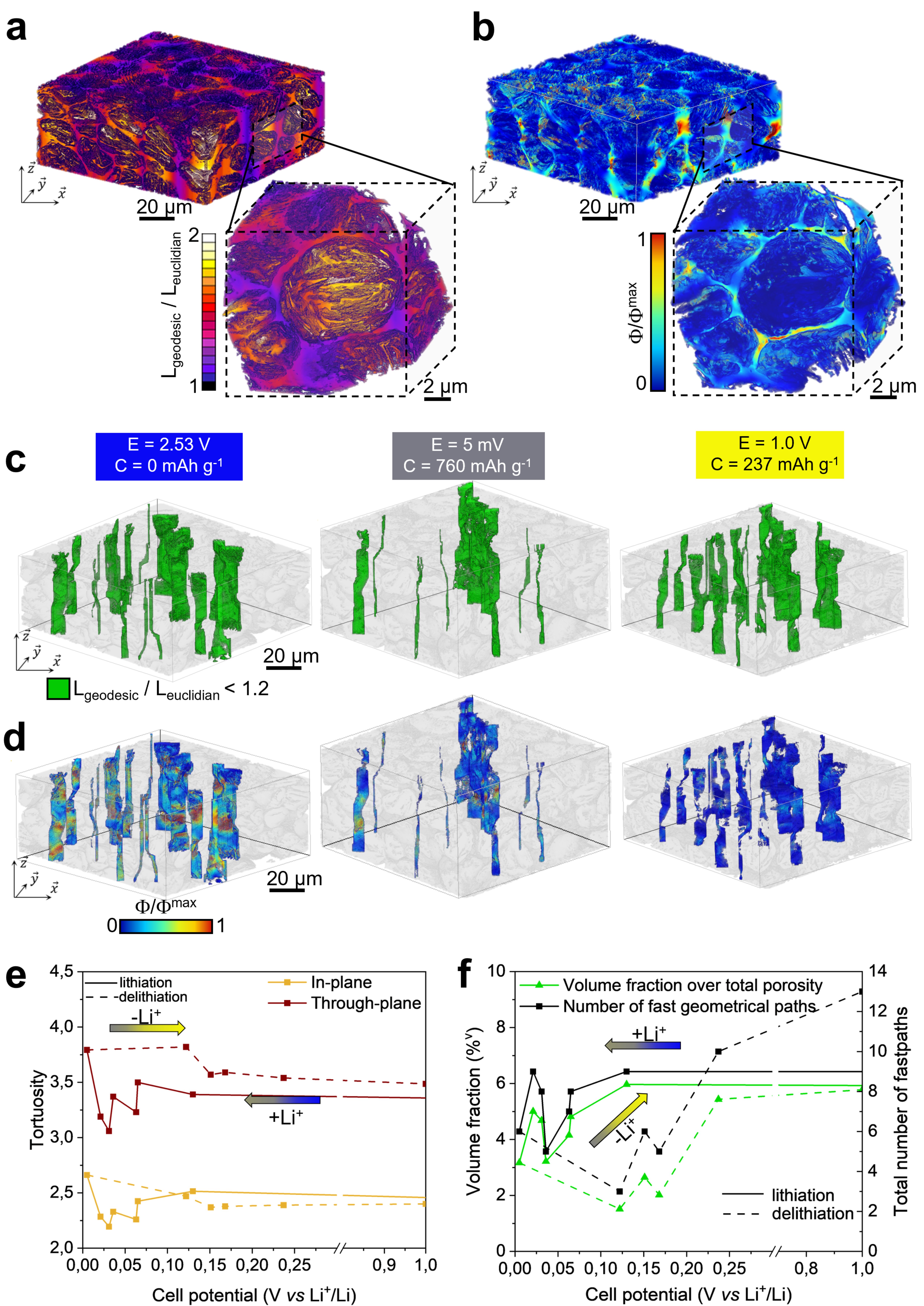}
  \caption{\textbf{Correlation between electrode deformation and ionic diffusion pathways.} 3D renderings of the a) through-plane geometric tortuosity ($\tau_{geo}$)  map of the Gr-Si composite electrode and b) flux map retrieved from solving the diffusion equation. Bottom: close up views highlighting the case of a chosen graphite particle along with its inner porosity. c) 3D renderings of the fast geometric paths ($L_{geodesic}$/$L_{Euclidean}$ $\leq$1.2 ) extracted from the geometric tortuosity map and d) corresponding Li-ions flux map along them at different steps during cycling. From left to right: initial state, end of \begin{math} 1^{st} \end{math} lithiation and end of \begin{math} 1^{st} \end{math} cycle. Evolution over the cell potential of e) the in-plane and through-plane tortuosity ($\tau$) estimated from solving the diffusion equation over the 3D the Gr-Si composite electrode volume. f) Volume fraction corresponding to the fast geometric paths over the total porosity and number of separated fast paths as a function of the cycling potential.}
  \label{tortuo}
\end{figure*}

The observed microstructural changes, including the building/release of local strained regions, are likely to alter the available pathways for ions to diffuse through the thickness of the electrode. The computation of the through-plane geodesic distance map in the reconstructed images enabled the quantitative assessment of the transport distances required for ions to travel from the separator to the current collector via the interconnected porous network of the electrode (Fig. \ref{tortuo}a). As highlighted in the 3D rendering obtained from the nano-CT scan acquired in the initial state (Fig. \ref{tortuo}a), the inner porosity of the graphite particles is well connected to the interparticle porosity, while it is associated with the longest geodesic paths (drawn in yellow in the figure). Using the same microstructural imaged data, flux maps were calculated by solving the diffusion equation at steady state. These maps represent the reduction in diffusive transport caused by the tortuous geometry of the medium (Fig. \ref{tortuo}b).\cite{COOPER2016203} The corresponding tortuosity value ($\tau$) associated to the effective diffusion coefficient was quantified (see Methods). Minima of Li-ions flux (in blue in the figure) are clearly identified inside the graphite particles porosity, where the geodesic paths are longer in average than in the inter-particles pores, whereas the maximum fluxes (in red) are distributed in the interparticle porosity. 

In the initial state, specific geodesic paths corresponding to geodesic distances less than 1.2 times the straight Euclidean ones were isolated in the structure (Fig. \ref{tortuo}c and Supplementary Fig. S8a). These features correspond to quasi-straight lines that cross the entire electrode thickness. They match with the highest Li-ions flux values and are denoted as fast geometric paths in the following. These fast paths are found right at the intersection with shorter geodesic paths in the in-plane direction (Fig. \ref{tortuo}d and Fig. S8b-e), showing that in-plane and through-plane mass transport phenomena are connected. 
During the formation cycle, both the number of fast geometrical paths and their corresponding volume fractions change, as seen in Fig. \ref{tortuo}c, showing that ionic diffusion paths are affected by the electrode deformation. This directly affects the values of in-plane and through plane tortuosity quantified in Fig. \ref{tortuo}e. Globally, the changes in tortuosity correlate to both the major rearrangement of the electrode microstructure evidenced in Fig. \ref{dvc_local}b,c) and diminution of the interparticle and intraparticle porosity seen in Fig. \ref{electrode_scale}c). In-plane values remain below 2.6, while through-plane values range between 3.3 and 3.8. They also correlate directly to the evolution of fast geometric paths, which reduce in number and volume (with respect to the total electrode porosity) during lithiation (Fig. \ref{tortuo}f, Supporting Video S5). This effect is not monotonic, but rather exhibits step-wise changes, especially near the activity voltage range of silicon, demonstrating how in-plane and through-plane microstructural rearrangements disrupt the ionic conduction paths. Later, during delithiation, the through-plane tortuosity value decreases rapidly after 150 mV \textit{vs.} $Li^{+}/Li$ and further stabilizes at the end of the formation cycle, reaching a value of 3.5, slightly higher than that of the initial state (Fig. \ref{tortuo}e, supporting Fig. S9). The in-plane component already shows a tortuosity reduction at the beginning of delithiation that can be associated with in-plane displacements when transitioning from lithiation to delithiation. Additionally, at the end of the cycling, the flux maps calculated on the fast geometric paths present low values ($\leq0.3$) and Li-ions fluxes distribution appear clearly more homogeneous (Fig. \ref{tortuo}d). These phenomena are correlated with the increased number of fast geometrical paths, from 9 initially to 13 at the end of the cycle (Fig. \ref{tortuo}f). We assume that the opening of new fast geometric paths leads to more regularly distributed Li-ion highways, acting towards a homogenization of the ions fluxes (Supporting Fig. S9). 

Overall, the mechanical irreversibility of the composite electrode leads to significant changes in the tortuosity and preferential ionic diffusion paths. Such distortion of the electrode microstructure during the formation cycle could result in a higher ability to deform in a systematic and reversible way during subsequent cycles, as also recently suggested by Valisammagari \textit{et al.}\cite{abi_dvc_2024} As mentioned earlier, the interparticle porosity is well connected to the inner porosity of the graphite particles, suggesting the importance of determining the chemomechanics at the particle scale to fully understand the couplings between electrochemical activity, deformations and mass transport channels.

\subsection{Chemomechanical evolution at the particle scale}\label{subsec6}
\begin{figure*}[h!]
\centering
  \includegraphics[width=10cm]{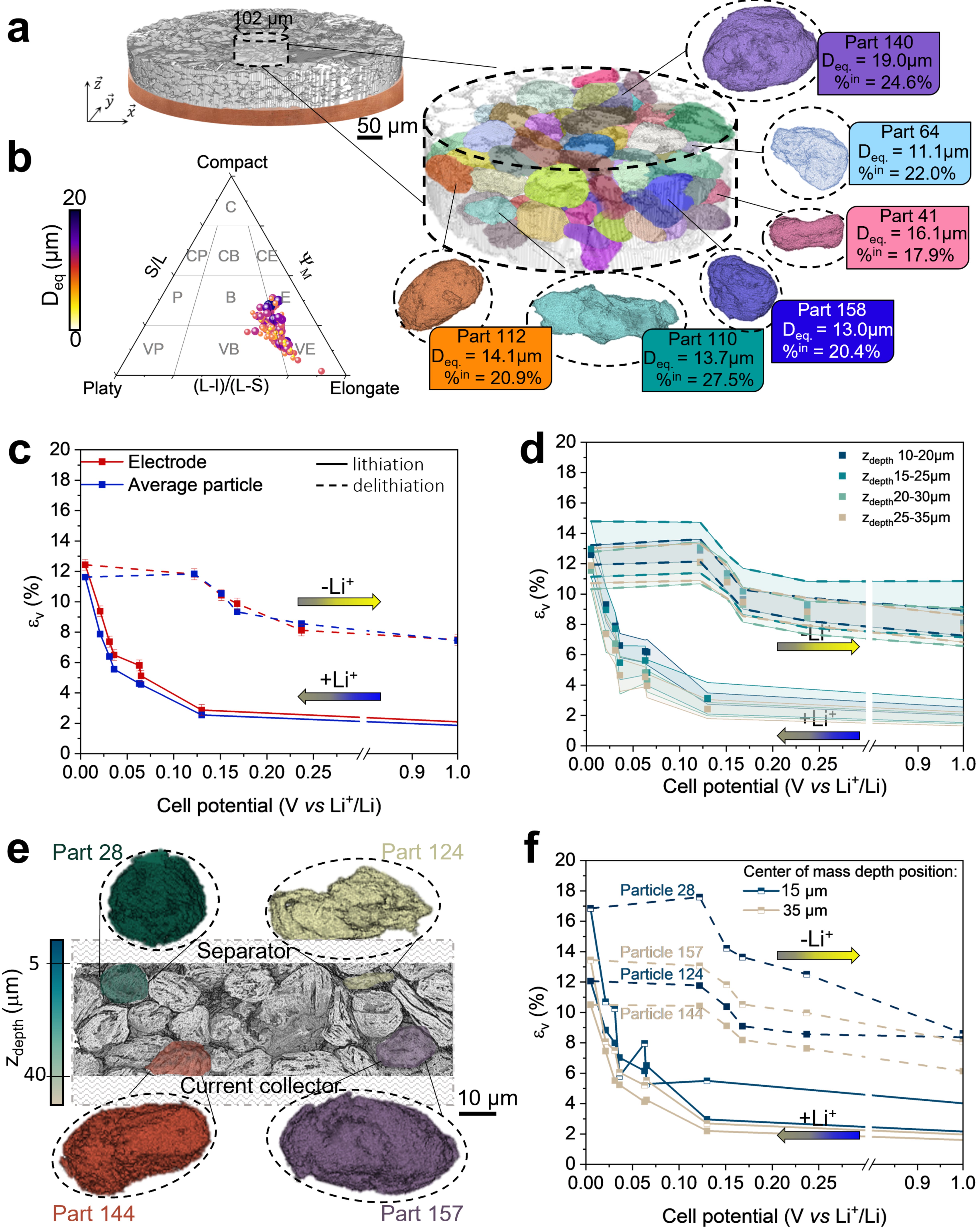}
  \caption{\textbf{Chemomechanics at the particle scale obtained by particle tracking.} a) 3D rendering of the Gr-Si electrode at low magnification (257 nm voxel size) along with a zoomed in view of the volume of interest for \textit{operando} acquisition (50nm voxel size) demonstrating the labeled graphite  particles selected for tracking (not touching the edges and corresponding to 45.5 \% of the total solid fraction in the imaged volume), with a selection of illustrative particles randomly distributed and their corresponding values of the equivalent diameter ($D_{eq}$) and the inner porosity volume fraction (${\%}^{in}$). b) Aspect ratio map (Sneed and Folk\cite{SneedFolk1958}) of the graphite particles showing their elongation (\textit{l/L}) compactness (\textit{S/L}) and projected sphericity ((\textit{${\psi}_{M}$})) (where \textit{S}, \textit{l} and \textit{L} are respectively the length of the shortest, intermediate and longest axis of the ellipsoid). The color scale corresponds to their equivalent diameter. c) Comparison of bulk electrode behavior through local DVC and averaged particle behavior through discrete DVC (particle tracking) for the volumetric strain evolutions during (de)lithiation over potential. d) Variability of the c) volumetric strain as a function of the particles' position across the electrode's depth ($Z_{depth}$, ranging from closer to the separator 10-20 ${\mu}$m to closer to the current collector 25-35 ${\mu}$m), during the formation cycle. The shadowed intervals highlight the standard deviation intervals for each $Z_{depth}$ range [$\overline\epsilon\pm\sigma$]. e) Cross section rendering with illustrative particles located closer to the separator (\texttt{\#}28 and \texttt{\#}124) and deeper in the bulk (\texttt{\#}144 and \texttt{\#}157), along with f) their corresponding volumetric strain evolution during the formation cycle.}
  \label{ddic}
\end{figure*}

Examining microstructural changes down to the particle scale is crucial to highlight the heterogeneities among individual particles and consequently understand their impact on the global composite electrode behavior. To achieve this, each graphite particle within a subset volume of the electrode ($z_{depth}$:5-40$\mu$m) was segmented by applying a sequence of image analysis operations (see Methods and supporting Fig. S10). From a total of 164 graphite grains, 66 were selected after excluding those touching the borders of the imaged volume, where information is missing and subsequent tracking would be error-prone. The selected particles account for 45.5 $\%^{v}$ of the total solid fraction in the imaged volume, as highlighted in the 3D rendering of Fig. \ref{ddic}a. 
Morphologically, they exhibit an average equivalent diameter of 11.3 ${\mu}$m and 24.7 $\%^{v}$ inner porosity (see Supporting Fig. S11a and S11b). Regarding their aspect ratio, these particles appear globally elongated regardless of their size, as illustrated in the Sneed and Folk graph of Fig. \ref{ddic}b (see Supplementary Note 5 for definition of S,l and L).\cite{SneedFolk1958}

Out of the 66 selected graphite particles, 56 (85 \%) were successfully tracked during the formation cycle through a so-called discrete DVC approach (see Methods) and the evolution of their respective grain-scale kinematics was measured. First, we verify that the mean volumetric strain of the individual particles matches the mean volumetric strain of the bulk electrode (Fig. \ref{ddic}c), showing consistency of the ensemble behavior determined by either bulk electrode-scale analysis or by averaging a representative set of individual particles. Then, to further investigate the influence of the local grain kinematics, the volumetric strain evolution of selected tracked particles is plotted as a function of their center of mass position in the depth of the electrode (\ie close to separator, middle of the electrode, or close to current collector). The particle volumetric strain reflects indirectly its degree of lithiation and is thus used here as a proxy to quantify the particle electrochemical activity. The average values (points) and associated standard deviations estimated within groups of 15 particles (shadowed areas) are shown in Fig. \ref{ddic}d for various electrode depth ranges. The complete dataset containing the values of all tracked particles is reported in the Supplementary Fig. S11a. 

Interestingly, the deformation behavior of the entire particle population shows lower variability during lithiation than delithiation.\cite{Sun:gy5070} On average, a trend of higher volumetric strain, \ie higher electrochemical activity, is observed closer to the separator, which agrees with previous reports discussing the balance between ionic versus electronic diffusion limited mechanisms.\cite{oney2025graphite, KisukKang2025, Pietsch2016_tomo_dvc, Eastwood2014} Note that the reported trend focuses on a 40${\mu}m$ deep section of the electrode, where a pronounced concentration gradient is present, as supported by complementary XRD mapping (see Supplementary Note 4 and Fig. S4a). However, this depth-dependent behavior does not uniformly hold when examining individual particle responses. Some particles located at similar depth in the electrode behave very differently. In the meantime, other particles behave similarly despite being at very different depths (Supplementary Fig. S12a). To illustrate this point, the behavior of particles located close to the separator (particles \texttt{\#}28 and \texttt{\#}124) is compared with particles deeper in the electrode bulk (\texttt{\#}144 and \texttt{\#}157) (Fig. \ref{ddic}e). Although they have approximately the same equivalent diameter and inner porosity (Fig. S11a, b) and are not attached to a large CBD-Si cluster, these particles differ in their particle/electrolyte surface fraction and proximity to fast geometric paths (Fig. S11c, d). In fact, at similar electrode depth, particle \texttt{\#}28 displays higher electrochemical activity than particle \texttt{\#}124, despite its lower particle/electrolyte surface fraction (64 \textit{vs.} 89\%). This enhanced activity can be attributed to its proximity to a fast geometric path, a condition similar to that shared by particle \texttt{\#}157, which, despite being located deeper in the bulk of the electrode, exhibits higher activity than particle \texttt{\#}124 positioned closer to the separator (Fig. \ref{ddic}f). 

These findings demonstrate the key role of the local particle environment and spatially distinct neighbourhoods, beyond bulk electrode microstructural features in the Gr-Si electrode. The proximity to fast geometric paths alters the driving force and equilibrium condition known to produce concentration asymmetry at interfaces due to ionic \textit{vs.} electronic driven diffusion limits. To fully capture the deviation of individual particles to the ensemble behavior, it is necessary to evaluate the impact of fast paths against other possible structural characteristics, as the vicinity to silicon clusters that can also drive local responses due to the large Li-ion fluxes and expansion/shrinkage of silicon which is likely to exert considerable local pressures on its immediate surroundings. 

\subsection{Influence of silicon clusters on graphite activity}\label{subsec7}
The link between local particle activity and global electrode behavior is now examined by studying the impact of the vicinity to CBD-Si on local deformations. Starting with the deviatoric strain evolution, the electrode level was compared with the particle average behavior. The mean deviatoric strain at the electrode level is notably higher, which is attributed to contributions coming from the large CBD-Si clusters and the grain-binder interfaces (white arrows in Fig. \ref{dvc_local}d, e and Supplementary Note 6, Fig. S13a). Also, no particular electrode depth-dependent trend is observed for the deviatoric strain, consistent with the fact that it depends on the grain-binder interfaces and large CBD-Si clusters that are randomly distributed along the electrode microstructure (Fig. S12b). 

To determine the local state-of-charge variability near the CBD-Si clusters, we decided to focus on the graphite activity in their surroundings. Throughout the electrode microstructure, subgroups of graphite particles were identified that are either tightly packed together or attached to a large CBD-Si cluster (Fig. S13b). In total, four large CBD-Si clusters (groups \texttt{\#}1,2 and 5,6, Supporting Fig. S13c,f,o,r) are observable in the imaged electrode microstructure, serving as particular regions of interest to follow the mechanical evolution of graphite particles attached to them during the formation cycle (see also supplementary Note 6). These groups are compared to groups of tightly packed graphite particles not attached to any CBD-Si cluster (group \texttt{\#}3 and \texttt{\#}4, Supporting Fig. S13i and l). 
High deviatoric strain accumulation is observed in some localized regions between graphite particles, as highlighted by the strain heatmap intensity in the reconstructed nano-CT images at full lithiation (Fig. S13c,f,i,l). This dilatant shearing behavior, typical of dense granular and/or cohesive materials, develops heterogeneously in the electrode microstructure. The motion correlation within and between these subgroups can be interdependent or even occurring in opposite directions, necessitating local microstructural rearrangements. Additionally, irreversible deviatoric strain is higher near large CBD-Si clusters, and can even reach unprecedented levels as displayed by the particles \texttt{\#}4 and \texttt{\#}138 of group\texttt{\#}1. These particles show large deviations with respect to the rest of the entire particle population, with higher volumetric strains at the end of first lithiation, especially after the potentiostatic hold at 5 mV \textit{vs.} $Li^{+}/Li$. As a consequence, these particles exhibit a high irreversibility (volumetric and deviatoric strains) at the end of the formation cycle, arising from the large volume changes associated with silicon delithiation which induce local rearrangements and shearing (Fig. S13d, e). Considering the macroscopic mechanical properties of the CMC binder from the literature,\cite{Li2007-yh} a local maximum tensile strain of 7\% or higher could lead to binder failure. Even though a different behavior could be expected at the microscale, this suggests that partial interface debonding from the conductive network is prone to occur when graphite particles are close to large CBD-Si clusters due to important tensile and shearing strains. Their volumetric strains are even increasing during the end of delithiation (up to 13\%), while they decrease for the rest of the particles (8\%). This could imply some partial lithium trapping by these graphite particles, as they possibly reached a lower potential after the disconnection event. 

Regarding the volumetric strain evolution, graphite particles attached to large CBD-Si clusters (Fig. S13d,g,p and s) tend to show slightly higher values compared to other subgroups not attached to clusters (Fig. S13j and m). The subgroup behaviors are also compared to the average population and its standard deviation ($\epsilon\pm\sigma$, represented in light gray in the two strain invariants graphs of Fig. S13). A more coherent behavior is observed for particles attached to the large clusters. Indeed, they exhibit lower deviations one with respect to another, as compared to other subgroups not attached to clusters, which in turn show slightly higher deviation ($\geq\sigma$) from the average behavior. This indicates potential structural stiffness in certain regions of the electrode where large CBD-Si clusters are present, as also suggested by Sun \textit{et al.}\cite{Sun:gy5070} However, the cluster-based observations must be taken with caution, as odd particles are also noticed in several groups, emphasizing the role of additional parameters in determining the local SoC status of the active material.

To summarize, we found that the local organization of the two active phases, one with respect to the other, is critical and influences the local chemomechanics. Graphite particles are most of the time influenced by the vicinity of silicon clusters that impact their local state-of-charge. Some individual particles clearly deviate from the ensemble, beyond statistical changes. These results indicate that there are multiple factors to account for predicting the ``best'' environment or condition favoring an efficient (hence, homogeneous) (de)lithiation, beyond the morphology at the electrode level or the typical ensemble metrics, such as global porosity or tortuosity. Specifically, couplings between local particle-scale features and global electrode characteristics play a decisive role in determining the particle chemomechanical activity in the Gr-Si system. Decorrelating the effect of each local parameter and classifying the main driving factors appear necessary to understand their relative importance and how they are intertwined.

\subsection{Key factors influencing chemomechanics}\label{subsec8}
\begin{figure*}[h!]
\centering
  \includegraphics[width=10cm]{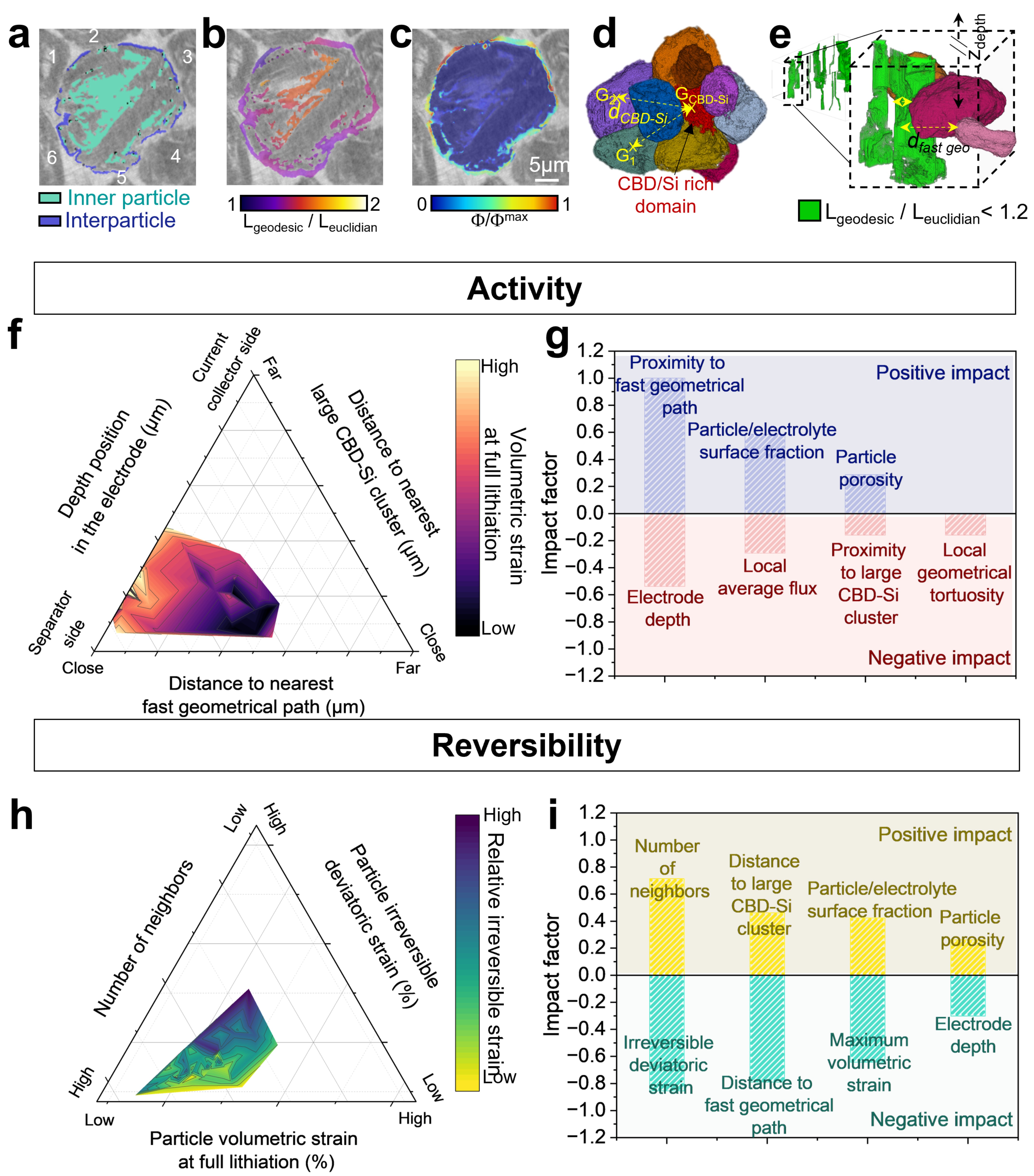}
  \caption{\textbf{Correlative  maps and key factors impacting electrochemical activity.} Local environment of the graphite particles showcasing the different parameters evaluated for the impact on the particle electrochemical activity, a) number of neighbors, electrolyte/particle surface fraction, particle inner porosity filled with electrolyte, b) average local geometric tortuosity, c) average Li-ions flux, d) Euclidean distance to nearest large CBD-Si cluster and e) nearest fast geometric paths. Ternary plots of the geometric environment impact and respective impact factors for the different parameters evaluated respectively on f, g) the electrochemical activity and h, i) relative irreversible behavior of the graphite particles.)}
  \label{corrmaps}
\end{figure*}

Chemical reactions in the Gr-Si electrode induce deformations that the microstructure has to accommodate. The local state of charge correlates to the local stress state and anisotropic displacement field induced by both structural constraints and by geometric factors. Across the electrode microstructure, particles exhibit different intrinsic properties, as well as local environment characteristics such as position, number of neighbors, \textit{etc}. Transient and/or irreversible phenomena are directly related to particle rearrangements and sliding within the cohesive medium formed by the active particles and the binder matrix, involving also changes in diffusion pathways, pore clogging and particle-binder debonding. We leveraged the statistical insights brought by the individual particle labeling combined with the discrete DVC approach to quantify the relative impact of an array of 14 local environment parameters on a particle's activity/irreversibility, categorized into four major types:
\begin{enumerate}
\item Particle morphology: equivalent diameter; volume; surface area; sphericity; electrolyte-filled inner porosity (light blue in Fig. \ref{corrmaps}a).
\item Diffusion properties: average geometrical tortuosity extracted from the global geodesic maps (L$_{geodesic}/L_{Euclidian}$, Fig. \ref{corrmaps}b); average Li-ion flux extracted from the global flux maps ($\Phi/\Phi_{max}$,Fig. \ref{corrmaps}c); particle surface in contact with the electrolyte normalized by the total particle surface, defined as ``particle/electrolyte surface fraction'' (Dark blue in Fig. \ref{corrmaps}a).
\item Geometrical positioning: depth position of the center of mass of the particle in the electrode ($Z_{Depth}$, Fig. \ref{corrmaps}e); Euclidean distance to the nearest large CBD-Si cluster ($d_{CBD-Si}$,Fig. \ref{corrmaps}d); Euclidean distance to the nearest fast geometrical path, among those which persist all along the formation cycle ($d_{fast}$, Fig. \ref{corrmaps}e); number of neighbouring graphite particles (5 in the example shown Fig. \ref{corrmaps}a).
\item Mechanical deformation: maximum volumetric and deviatoric strain values at full lithiation ($\epsilon_{VLith}, \epsilon_{DLith}$) ; irreversible deviatoric strain at the end of the formation cycle ($\epsilon_{VdeLith}, \epsilon_{DdeLith}$) .
\end{enumerate}

To quantify the relative influence of each parameter, the activity of a selected particle was considered to be proportional to the volumetric strain at full lithiation, and plotted using parameter-sensitive graphs (Fig. \ref{corrmaps}f). We demonstrate that the most electrochemically active particles, exhibiting the highest volumetric strain, are located rather close to fast geometrical paths and near the separator/electrode interface.
Regarding the distance from large CBD-Si clusters, the particles attached to them are more prone to experience larger volumetric strain due to the Si particles and their lithiation reaction that could drive a higher flow of charges, enhancing local ionic and electronic fluxes. But, the further graphite particles are located from these clusters, first their electrochemical activity will drop slightly, then increase back due to the overcoming contributions of other local environment parameters. To evaluate whether correlation between the electrochemical activity and the evaluated parameter is strong, the degree of correlation was estimated through a Pearson’s correlation (see Methods). Additionally, a metric was introduced to quantify the impact of a given parameter (for instance, the distance to nearest fast diffusion path) on the activity. This impact factor ranges between 0 (no influence) and 1 (strong influence) (see Methods and Supplementary Note 7, Fig. S14). Among all the 14 evaluated parameters, nine of them exhibit correlation factors higher than 0.35, indicating a strong relationship with electrochemical activity and/or irreversibility, but not exclusively. Based on all the studied local environment parameters provided by the \textit{operando} nano-CT, up to 90\% of the standard deviation of the particles from the ensemble behavior can be rationalized in the Gr-Si composite electrode, compared to only 45\% when restricted to commonly used macro-scale parameters. In particular, for electrochemical activity, seven parameters are reported, which are ordered by degree of importance in Fig. \ref{corrmaps}g, namely: distance to fast geometrical paths, particle/electrolyte surface fraction, particle inner porosity, average local geometrical tortuosity and Li-ions flux, depth position in the electrode, distance to large CBD-Si clusters. Note that the correlation factor values are detailed in the supporting information. 

Looking at the impact factor, positive or negative, of the various parameters on the electrochemical activity in Fig. \ref{corrmaps}g, we see that the proximity to stable fast geometric diffusion paths, the particle/electrolyte surface fraction, and the electrode depth have the strongest influence. The first two characteristics fall in the positive (blue) impact area, meaning that the higher these parameters, the higher the expected positive impact on the electrochemical activity. Conversely, the impact of the parameters in the red region is negative, meaning that they need to be reversely and effectively adapted to increase the electrochemical activity. Notably, within the top 33\% of the electrochemically most active graphite particles, 70\% are connected to constant fast geometric paths, while the remaining fraction comprises particles near the separator with high particle/electrolyte surface fractions. Although the electrode depth has often been identified as a key parameter driving lithiation heterogeneity - an observation which also holds true here when averaging the local behaviour of multiple particles (Fig. \ref{dvc_local}d) - the high resolution particle-by-particle analysis presented in this study reveals additional parameters of critical importance. Specifically, proximity to large CBD-Si clusters is found to be detrimental to high electrochemical activity when not directly in contact (Fig. \ref{corrmaps}f), alongside with important local geometrical tortuosity and/or high average Li-ion flux. 

The same procedure was applied to assess particle irreversibility, which was quantified using the relative irreversible volumetric strain (Fig. \ref{corrmaps}h). Based on the correlation results, we identify novel key factors influencing the irreversible behaviour of the composite electrode. In addition to the parameters listed for the electrochemical activity, three dominant parameters are found, namely: the number of neighboring particles, the maximum volumetric strain and irreversible deviatoric strain (Fig. \ref{corrmaps}i). For example, regions exhibiting significant irreversible deviatoric strain correspond to zones of high local volumetric irreversibility, which correlate with proximity to large CBD-Si clusters and/or to the separator (Fig.\ref{electrode_scale}e and Fig. S13c,f). Distance from fast geometric paths also critically affects irreversibility, with particles located further away delithiating more slowly and consequently more prone to experience irreversible strain at a fixed cut-off voltage. Interestingly, a higher number of neighbouring graphite particles appears to preserve the microstructural integrity of the electrode and reduce local irreversibility, likely due to the constraining effect of locally compacted areas with numerous contact points leading to a less significant deformation. In terms of graphite particle reversibility, the inner porosity can have a positive impact as well, while higher fraction of intra-pores could accommodate the particle volume change more easily through elastic energy/stress dissipation. 

\section{Discussion}\label{sec3}
This work demonstrated a customized \textit{operando} X-ray nano-holo-tomography workflow, based on an in-house developed miniaturized cell and setup, that enabled tracking of the morphological evolution of a Gr-Si composite electrode during its formation cycle in a reliable, reproducible and standardized way at high spatial resolution. Such meticulously controlled experimental conditions combined with the highly coherent synchrotron nanobeam were needed to gather real-time data on a cycling anode with unprecedented morphological details. The link between microstructure and Gr-Si electrode performance was revealed by combining electrode-level analysis in terms of porosity and particle volume fraction, with particle-by-particle tracking and quantification of local states of charge and stress. Our method is suited to identify and classify the many competing structural factors that regulate electrochemical activity and mechanical reversibility. In the Gr-Si system, we showed that (i) the electrode expansion/contraction along with its porosity evolution follows a multi-step process during (de)lithiation, where the overall microstructural expansion (contraction) is followed by an abrupt decrease (increase) of the porosity, primarily driven by silicon activity, consistent with mesoscale results obtained on bulk Gr-Si electrodes \cite{Berhaut2023}; (ii) the electrode deformation is highly anisotropic and irreversible, as well documented in the literature on silicon anodes,\cite{Chen2025} and it is dominated by the deformation of the graphite particles, demonstrating their role as the backbone structure in the composite electrode. Higher silicon content electrodes, which might be needed in the future for high energy density Li-ion batteries \cite{Chen2025}, could face a difficulty in this regard, as the role of graphite would be weakened and additional mechanical constraints could destabilize the composite architecture, ultimately reducing cyclability\cite{Moon2021-iz, GasteigerJES2017-vu}; (iii) the exhibited morphological changes lead to important rearrangements that affect the ionic conduction paths, in particular fast geometrical paths enhancing local electrochemical activity; (iv) the spatial distribution of large CBD-Si clusters in the electrode microstructure locally affects the activity/reversibility of surrounding graphite particles; (v) at the particle level the key parameters beneficial to local electrochemical activity are the proximity to fast diffusion paths, the particle/electrolyte coverage, and the particle inner porosity, while the ones impeding it are the depth within the electrode, the local average Li-ions flux, and the proximity to large CBD-Si cluster; (vi) in terms of mechanical reversibility, the main contributing factors are the number of neighboring particles, the distance to large CBD-Si cluster, and the particle/electrolyte coverage; in contrast, important irreversible deviatoric strain and maximum volumetric strain at full lithiation lead to increased mechanical irreversibility.

Incorporating these ingredients in modeling would be crucial to predict the electrode behavior and build effective digital twins accounting for the 3D complexity of real microstructures. Common approaches use ensemble and space-averaged descriptors to describe drying or calendaring steps,\cite{Franco2019, Franco2021} while few works use real microstructures as inputs of simulations.\cite{Lu2020, ZChen2017_LMnO, Yan2012-lk} For instance, lithium and stress distributions were captured in commercial $LiFePO_{4}$ electrodes with heterogeneous microstructures using 3D particle-resolved model,\cite{Mai2019-lh} while inhomogeneous spatial distribution of carbon binder was probed using 3D reconstructed model built from tomographic data.\cite{Hein2020-lp} However, to the best of our knowledge, there is no detailed modeling attempted using data from \textit{operando} nano-CT on battery materials, a gap that should be filled to advance our understanding and boost modern reverse engineering that could be obtained if models could be trained using selected features such as the fast ion diffusion paths or the collection of local features ranked using our impact scale. We can anticipate that the advent of experimental datasets as precise and reliable as the one reported here will accelerate the digital development of homogeneous-by-design ionic medium, meeting the stringent requirements of future battery materials for Li-ion and beyond technologies. 

Moreover, our findings can serve to tailor the electrochemical activity/reversibility by providing practical directions to reversely improve the electrode manufacturing and processing, beyond ensemble parameters: 
\begin{itemize}
    \item Electrode engineering should promote the existence and stability of fast geometric diffusion paths across the electrode microstructure. Some attempts in this direction were made by using magnetic field-induced preferential orientation of graphite particles \cite{Billaud2016-ug, Zhang2019-az}, although performance improvement was not demonstrated. Templating and nanostructuring approaches were also explored widely in silicon-based materials to design preferential 1D or 2D ionic channels,\cite{Wada2014-me, Maroni2022-yz} as well in other systems as oxide thin films\cite{Han2023-cl} or solid state electrolytes.\cite{Weng2023}
    \item Implementation of a bimodal particle size distribution could achieve increased packing density, thereby improving electrode compactness and enhancing mechanical reversibility. This can also benefit the electrochemical performance thanks to a potentially more uniform local current distribution, as suggested by results from pseudo two-dimensional model (P2D).\cite{Taleghani2017-md}
    \item Mitigation of the heterogeneities in the composite microstructure can be achieved by ensuring a more homogeneous distribution of Si particles, leading to a more uniform expansion/contraction and hence higher reversibility. This can be directly addressed by optimizing the slurry homogeneity together with the subsequent step of electrode drying process, but also addressed by materials engineering.\cite{He2025-yz} It is worth noting that this task is becoming even more critical with the potential use of silicon-based anolytes for all solid state batteries, where the mixing of silicon architectured materials with argyrodites or halides must be optimized to ensure interfacial stability and long-term cyclability,\cite{Ahmed2025-gi, Nelson2024-dk} requiring control over intimate phase-distribution using suited milling procedures.
\end{itemize}

The \textit{operando} nano-CT characterization is a powerful tool applicable across multiple length scales, and potentially extendable to a wide range of energy conversion and storage devices. By bridging the electrode and particle scale dynamics, the origin of the local chemomechanical heterogeneity throughout the electrode was revealed. We confirmed that, on average, the particles electrochemical activity tends to decrease when they are located closer to the current collector interface, a well-known phenomenon ascribed to diffusion limited processes amplified at high C-rates or in thick electrode designs, potentially leading to lithium plating or other degradation mechanisms. Nevertheless, we found that this effect can be counterbalanced by local variations in particle-to-particle behavior, which can play a prominent role. We established a comprehensive relationship between local electrode deformation and specific ionic diffusion pathways that rearrange during the electrode cycling, discovering in particular the presence of fast geometrical paths that affect the global Li-ions flux distribution and local electrochemical activity. 
Moreover, subgroups of graphite particles deform differently from the global electrode, depending on their relative position with respect to local large CBD-Si clusters. As a consequence, the electrode’s morphological evolution is highly heterogeneous, with regions characterized by local particle movements, ultimately driving localized microstructural rearrangements during cycling. This implies that the electrode state cannot be determined by considering only its intrinsic characteristics, such as volume fractions, loading, particle sizes, macroscopic porosity and tortuosity, nor by the precise knowledge of the operating conditions (voltage, temperature, pressure, C-rate, etc.): additional criteria must be urgently considered for their practical and digital manufacturing. Similar local-scale phenomena were identified in garnet-type solid electrolytes where grain-based information was mapped, resulting in correlations of chemomechanics with bulk microstructure.\cite{Dixit2022} Hot spots of mechanical and ionic constrictions were found, and considered as nucleation spots for degradation, leading to the conclusion that failure could be a stochastic process affected by local microstructural heterogeneities. Hence, determining the main driving forces that regulate mass transport and charge transfer mechanisms, ultimately driving system performance, requires to break down the observation scales, accounting for the complex granular nature of composite battery components.

\section{Methods}\label{sec4}

\subsection*{\textit{Operando cell}}\label{subsec1}
An in-house miniaturized cell was specifically developed for \textit{operando} nano-CT. It is made up of two electrodes placed on both sides of a glass fiber separator soaked in the electrolyte. Stainless steel connectors terminated with a thread ensure cell compression and proper sealing with a constant pressure (23 kPa estimated with a dynamometric wrench). Note that the upper connector was designed shorter in order to minimize its weight and attenuate cell wobbling during tomography acquisition. The cell body is made of perfluoroalkoxy polymer (PFA), which is transparent to both X-rays and visible light for proper assembly in the glovebox and reliable tomography measurement. Additional sealing is performed at the junction between the cell body and the connectors by means of UV-polymerizing glue (Krylex). More details concerning the cell dimensions and setup are available in  Vanpeene \textit{et al.}\cite{Vanpeene2025}

\subsection*{Sample preparation}\label{subsec2}
The Gr-Si raw electrodes were produced by CIDETEC in the framework of the European project BIGMAP.\cite{BIGMAP} Their composition is graphite (74 wt.\%, from Imerys), nano Si@C (11 wt.\%), SBR:CMC (1:1) (14 \%wt, from SBR Zeon BM451B 40\%, CMC Chempoint W2000 3\%) and Super P (1 wt.\%, from Timcal®) as conductive additive. The electrode capacity is 3.3 mAh $cm^{-2}$, \ie 670 mAh $g^{-1}$.
The samples used for \textit{ex situ} and \textit{in situ} measurements were prepared by femto laser cut of the raw electrode films thanks to a Speedy 300 laser engraving system from Trotec to achieve a final diameter of 0.7 mm.
Half-cells were assembled in an Ar-filled glovebox ($O_{2}$ $\leq$ 0.5 p.p.m.) and composed of a lithium metal disc (1 mm diameter) placed on top of a precut 1 mm Cu current collector (thickness = 240 ${\mu}$m) on one side, and the Gr-Si electrode cut by femto laser on the other side. Two layers of glass microfiber were used as separator (grade GF/D; diameter = 2 mm; thickness = 670 ${\mu}$m; Whatman) and soaked with 10 ${\mu}$L electrolyte made of ${LiPF_{6}}$ 1M in EC:EMC (3:7) + 10 wt.\% FEC (Solvionics). In addition, coin cells were also assembled in full cell configuration using LNO (1.0 mAh $cm^{-2}$; BASF) and Si-Graphite (1.11 mAh $cm^{-2}$; Cidetec) electrodes and 50 ${\mu}$L of electrolyte following a standardized protocol for cell assembly. Cells were cycled galvanostatically for 100 cycles after 6 h of rest and formation step composed of 3 CCCV charge and CC discharge at C/10 between 2.5 - 4.2 V (see Fig. S15). 

\subsection*{Data acquisition and electrochemistry}\label{subsec3}
The electrodes were cycled in galvanostatic mode at full capacity between 5 mV and  1 V \textit{vs.} $Li^{+}/Li$ at a current density of 110 mA $g^{-1}$ (C/6) in (de)lithiation using an OrigaFlex OGF500 potentiostat/galvanostat from OrigaLys. A potentiostatic hold was added after reaching the cut-off voltage at the end of lithiation at 5 mV \textit{vs.} $Li^{+}/Li$. Tomography acquisitions were performed every hour or when a specific voltage limit was reached. During measurement time, the \textit{operando} cell relaxed to open circuit voltage, which corresponded to a total shift of approximately 50 mV over a period of 12 min (Fig. S5a). The cell voltage retrieved its previous values within a minute when switching back the current.

X-ray nano-tomography acquisition was performed at the ID16B beamline of the ESRF,\cite{Martinez-Criado:hf5304} using the holo-tomography technique \cite{holo_cloetens_1999} with a conic beam produced by a nano-focused beam ($H: 55 nm  \times V: 60 nm$) allowing multiscale measurements thanks to geometrical magnification. Low magnification acquisition of the Gr-Si composite electrode performed at 257 nm voxel size (field of view ${526 \times 526}{µm^{2}}$, 1 min 30s acquisition time) coupled with higher resolution \textit{ex-situ} scans at 50 and 25 nm,voxel sizes (field of view respectively of ${102 \times 102}{µm^{2}}$ and ${51 \times 51}{µm^{2}}$, 12 min acquisition time each). 3203 projections with 35 ms exposure time for all different pixel sizes along a 360 deg. rotation, as well as 20 and 21 reference and dark images respectively, were recorded on a PCO edge 5.5 CMOS camera (${2048 \times 2048} pixels^{2}$) equipped with a 30 ${\mu}$m thick LSO scintillator. Repeated acquisitions were performed at four different sample-to-detector distances, in order to reconstruct a full phase image, \cite{Zabler2005, holo_cloetens_1999} providing high contrast at high resolution. For the incident X-ray beam, the energy was 29.4 keV and photon flux $3.4\times10^{11}$$ph s^{-1}$.

The multiscale acquisition scheme allowed for a quantitative description of the complex composite electrode microstructure, while ensuring the representativeness of the imaged volume (Supplememtary Fig. S1,S2). The determination of the minimum representative elementary volume is detailed in the Supplementary Note 1 and in a previous work.\cite{Vanpeene2025} The ground principle aims at calculating the standard deviation of the material volume fraction compared to the theoretical value in sub-volumes of different sizes ranging from 22 to 250 ${\mu}$m for incremental steps of 25 ${\mu}$m.

For the \textit{operando} measurement, the sample was placed 52 mm away from the focal spot (${55 \times 60}{nm^{2}}$) and 280 mm from the detector, leading to a final voxel size of 50 nm and a field of view of ${102 \times 102}{{\mu}m}^{2}$. Keeping the same set of acquisition parameters as the \textit{ex-situ} measurements, the number of projections was reduced to 2205 to reach higher temporal resolution with 9 min duration for a full X-ray nano-holo-tomography scan at 50 nm.

3D reconstructions were achieved in two steps: (i) phase retrieval calculation using an in-house developed octave script based on a Paganin-like approach using a delta/beta ratio of 2744, and (ii) filtered backprojection reconstruction using ESRF software PyHST2.\cite{MIRONE201441} Final volumes of ${526 \times 526\times 526}{{\mu}m} ^{3}$ ${102 \times 102\times 102}{{\mu}m} ^{3}$ and ${51 \times 51\times 51}{{\mu}m} ^{3}$ with respective voxel sizes of 257, 50 and 25 nm  were obtained in a 32-bit floating point. Post-processing ring removal and 16-bit conversion was performed using a dedicated Matlab script.\cite{lyckegaard2011correction}

\subsection*{Image analysis}\label{subsec4}
Different sub-volumes, extracted from the entire reconstructed volume (${102 \times 102\times 102}{{\mu}m}^{3}$), were used for the different image analysis steps of the \textit{operando} images. Note that the electrode itself corresponds to a 52 ${\mu}m$ height within this volume. 

\begin{itemize}
    \item Image segmentation: Due to the important phase shift at the interface with the current collector, leading to imaging artefacts in the respective positions of the reconstructed volume, the phase segmentation step was performed on a sub-volume of ${102 \times 102\times 40}{{\mu}m}^{3}$ within the electrode. The different phases were segmented using a machine learning based random forest classifier embedded within the Ilastik software.\cite{Berg2019-xu} The corresponding volume fraction of graphite, porosity (inner and inter-particle) were computed as represented in Fig. \ref{electrode_scale}a-c.
    \item Large CBD-Si cluster: To estimate the volume change of the CBD-Si rich clusters depicted in Fig. \ref{electrode_scale}d-f, three sub-volumes of ${20 \times 20\times 13}{{\mu}m}^{3}$ were considered and averaged. The analyses were performed on the reconstructed 16-bit images using the ImageJ software.\cite{Schindelin2012} Segmentation in these regions was achieved using a machine learning based random forest classifier embedded within the Ilastik software.\cite{Berg2019-xu}
    \item DVC: The same sub-volume of 40 ${\mu}m$ height as for the segmentation step was used for the local DVC calculation. More details on the procedure are listed hereafter.   
    \item Tortuosity estimation: A parallelepiped sub-volume of ${62 \times 62\times 40}{{\mu}m}^{3}$ within the electrode was required for this step accounting for the diffusion boundary condition problem. The geodesic distance maps were calculated through an in-house developed script based on previous work of Chen-Wiegart \textit{et al.}\cite{CHENWIEGART2014349} implemented in ImageJ. The tortuosity estimation were performed using the TauFactor matlab package.\cite{COOPER2016203} More details on the procedure are listed hereafter.
    \item Particle labeling and discrete DVC: Graphite particles within a depth of 5 to 40${{\mu}m}$ inside the electrode were labeled on a circular section of ${102 \times 102}{{\mu}m}^{2}$. Particle parameters (center of mass position, marching cube volume/surface, sphericity, equivalent diameter, inner porosity fraction, number of neighbors, Euclidean distances to nearest fast geometrical path and large cluster of CBD-Si, local Li-ions flux and geometrical tortuosity) were calculated through an in-house developed script implemented in ImageJ. 
\end{itemize}

\subsection*{Tortuosity calculation}\label{subsec5}
The tortuosity calculations were performed solving the diffusion equation:
\[ D_{eff} = D.\frac{\epsilon}{\tau} \]

where $\epsilon$ is the volume fraction of the ionic conductive phase; \textit{D} its intrinsic diffusivity; and \textit{$D_{eff}$} its effective diffusivity through a porous volume. The tortuosity factor ($\tau$) was obtained from a simulation while comparing the steady-state diffusive flow through a pore network with that through a fully dense control volume of the same size.\cite{COOPER2016203} This steady-state diffusion problem satisfies the following set of equations, including the fixed Dirichlet conditions imposed at the two parallel boundaries identified thanks to their normal vector \textit{n}:

  \[\nabla^2{C} = 0 \]
  \[\nabla{C}.n = 0 \]
  \[C_{collector} = 0 \]
  \[C_{separator} = 1 \]  

In parallel, the geometrical tortuosity was estimated from the geodesic distance map as the ratio of the shortest length path from point A to point B over the corresponding Euclidean distance. 
\[ {\tau}_{geo} = \frac{L_{geodesic}}{L_{Euclidean}} \]

\subsubsection*{Graphite particle labeling}\label{subsec6}
A customised graphite particle labeling procedure in \texttt{Python} was specifically designed here, due to the limitations of traditional watershed algorithms to separate particles presenting large internal porosity that are, typically for graphite, folded and pressed against each other. The different steps are summarized in the supporting information (Fig. S10). Starting from the binarised segmented image of the particles at the initial state, a first step of internal porosity clogging was performed using a set of max filters (2 voxels distance for 3 iterations) to dilate the active material phase. Then, the remaining phase outside the active material region was used as a starting level-set to back-propagate boundaries in order to capture the external shape of the particles thanks to a morphological operation derived from the Chan-Vese method (\href{https://scikit-image.org/docs/stable/api/skimage.segmentation.html#skimage.segmentation.morphological_chan_vese}{sckimage Chan-Vese}) available within the \texttt{scikit-image} library,\cite{scikit-image} controlling the number of iterations, region homogeneity indices and level-set smoothness. Note that this step allowed the recovery of some of the finer particle separation channels without recovering too much of the internal porosity.  This resulting filled-in graphite phase was used to compute an internal distance map from its borders. Subsequently, a watershed seeding was performed based on the local maxima in the obtained distance map. Local maxima were filtered according to a minimal threshold value (linked to the smallest particle dimension) and the separation distance between each seed (linked to the particle average characteristic dimension), giving priority locally to the highest values. Finally, watershed propagation was performed within the active material phase based on the previously computed seeds.

\subsection*{Statistical Correlation}
Different parameters \textit{p} were investigated for the particle \textit{i} local environment correlation with its electrochemical activity $X_{ac}^{i}$. Among the population of particles,  $X_{ac}^{i}$ fluctuates between a minimum $X_{ac}^{min}$ and maximum $X_{ac}^{max}$ value, with a mean value of $\overline{X_{ac}}$. For each evaluated parameter \textit{p}, the average value $\overline{p}$ was estimated for each $p^{i}$ value measured among the particle population. The correlation factor between the electrochemical activity $X_{ac}$ and each parameter \textit{p} was then defined as:
\begin{equation}
    corr(X_{ac}, p)= \frac{cov(X_{ac}, p)}{\sigma_{X_{ac}}.\sigma_{p}}
\end{equation}
Thus for n measurements of ($X_{ac}^{i}$, $p^{i}$):
\begin{equation}
    corr(X_{ac}, p)= \frac{\sum_{i=1}^{n} (X_{ac}^{i}-\overline{X_{ac}}).(p^{i}-\overline{p})}{\sqrt{\sum_{i=1}^{n} (X_{ac}^{i}-\overline{X_{ac}})^{2}.\sum_{i=1}^{n} (p^{i}-\overline{p})^{2}}}
\end{equation}

The unexplained fraction of the deviation from the ensemble behavior is computed as the sum of squares of residuals-that is, the sum of squares of the prediction errors-divided by the sum of squares of deviation of the values of the dependent variable from its expected value. In order to quantitatively compare the influence of the different tested local environment parameters \textit{p}, we introduced the metric of impact factor, defined as the normalized slope between two selected points ($X_{ac}^{i}$, $p^{i}$) and ($X_{ac}^{j}$, $p^{j}$) such as:
\begin{equation}
  IF= \displaystyle\left\lvert\frac{\frac{X_{ac}^{i}}{X_{ac}^{max}}-\frac{X_{ac}^{j}}{X_{ac}^{min}}}{\frac{p^{i}}{p^{max}}-\frac{p^{j}}{p^{min}}} \right\rvert
\end{equation}

for a parameter \textit{p} in [$p^{min}$; $p^{max}$] and an electrochemical activity $X_{ac}$ in [$X_{ac}^{min}$; $X_{ac}^{max}$]. The impact factor ranges between 0 and 1. More information are detailed in Supplementary Note 7.

\subsection*{Digital Volume Correlation}
The open-source software \texttt{SPAM} \cite{Stamati2020} was used to perform Digital Volume Correlation analysis and compute the kinematic fields during the \textit{operando} test.
The correlation procedure in \texttt{SPAM} aims to measure a linear and homogeneous deformation function, $\F$, such that a material point located at $\x$ in the reference image corresponds to the same material point at $\x' = \F\cdot\x$ in the deformed image.
This deformation function $\F$ is represented as a $4\times4$ matrix that contains $12$ unknowns, accounting for affine transformations: translation, rotation, normal and shear strain.
The formulation of the correlation algorithm is based on a gradient-based iterative procedure that minimises the difference between the reference and the deformed image, progressively correcting the latter by a trial deformation function (see \cite{Tudisco_2017, stamati:tel-02923399} for more details).
Convergence is assessed by the norm of the increment of the deformation function between two successive iteration steps.

A total DVC analysis was conducted by mapping the initial imaged volume at fresh state (\begin{math} 102\times102\times42 {{\mu}m}^{3} \end{math}) with each subsequent volume scanned during the first cycle, as opposed to an incremental analysis which maps a pair of consecutive electrochemical states.
For the multiscale analysis, two distinct DVC strategies were applied: a local/continuous approach at the electrode level and a discrete approach at the particle level.
In both cases, the initial step was the alignment of each pair of images through a rigid-body registration measured at the centre of mass of the electrode (capturing overall displacements and rotations), which was then subtracted from the local fields measured at either the bulk or particle scale.

The local approach at the electrode bulk consisted of defining a set of nodes in the reference image (fresh state) around which independent cubic subvolumes (\ie correlation windows) were extracted and searched in the deformed images (at different cycling states) using the iterative algorithm described earlier.
This process yielded, for each correlated pair, a deformation function $\F$ at the center of each window, resulting in a 3D field of deformation measurements across the entire reference image. 
A pyramidal coarse-to-fine scheme was applied, involving for each pair of images, successive computations with gradually smaller correlation window sizes $(20,~8,~4)~\mu\text{m}^3$, keeping a constant $50\%$ overlap between neighbouring windows.

The finest pyramid level $(4\times4\times4~\mu\text{m}^3)$ resulted in the measurement of a heterogeneous deformation field of high spatial resolution (with many nodes falling well inside the graphite grains and their respective interfaces) and was selected based on a DVC uncertainty study detailed in Supplementary Note 8 and Fig. S16,S17. For each correlated pair, the displacement field was obtained by extracting the translation component of each $\F$. To compute the strain fields, a linear mapping of $2\times2\times2$ neighbouring displacement measurements on Q8 shape functions was applied producing a field of transformation gradient tensors $\boldsymbol{F}$. Following the finite large-strain framework, a polar decomposition of $\boldsymbol{F}$ for each Q8 element, yielded the right stretch tensor $\boldsymbol{U}$ and the rotation tensor $\boldsymbol{R}$.
The volumetric strain was determined by the Jacobian determinant of $\boldsymbol{F}$: 

\begin{equation}
\varepsilon_\text{V} = det(\boldsymbol{F})-1
\end{equation}

Based on a multiplicative decomposition of $\boldsymbol{U}$ into an isotropic and deviatoric part, the deviatoric strain was determined by the norm of the deviatoric tensor: 

\begin{equation}
\boldsymbol{U}^\text{dev} = det(\boldsymbol{F})^{\frac{-1}{3}}\boldsymbol{U}; \quad\varepsilon_\text{D} = ||\boldsymbol{U}^\text{dev}-\boldsymbol{I}||
\end{equation}

For the correlation at the particle scale, a discrete approach was applied by tracking the labeled graphite particles throughout the formation cycle.
This involved defining arbitrarily shaped correlation windows in the form of subvolumes (\ie bounding boxes) that enclosed each labeled particle in the fresh state, centerd on the particle’s center of mass.
The grayvalues inside each bounding box were then extracted and matched with the deformed images (at different cycling states) based on the iterative algorithm mentioned previously.
A deformation function $\F$ was thus measured at the center of mass of each particle, producing for each correlated pair, a second 3D deformation field, this time defined at the center of mass of the particles at the fresh state.
The strain invariants (volumetric and deviatoric) for each particle were directly calculated through a polar decomposition of the transformation gradient $\boldsymbol{F}$ extracted from the measured $\F$.
Similarly with the local approach, a DVC uncertainty analysis was conducted for the discrete correlations, detailed in Supplementary Note 9 and Fig. S16,S18.

\backmatter

\bmhead{Supplementary information}
The supplementary file is available for complementary information about representativeness, electrochemical data and beam damage assessment, segmentation and labelling technique, digital volume correlation and statistical analysis. See "Supporting Information: 4D operando X-ray nano-holo-tomography reveals multiscale chemomechanics in Silicon-Graphite anodes.pdf".

\bmhead{Acknowledgements}
The authors thank the European Synchrotron Radiation Facility for provision of beamtime for the measurements at the ID16B beamline in the frame of the long-term proposal MA4929 "Multi-scale Multi-techniques investigations of Li-ion batteries: towards a European Battery Hub". This work has been supported by the Science Impulse program, Institut Carnot, DRT-CEA, project "Big data for better lithium-ion battery characterization". This work has received funding from the European Union’s Horizon 2020 research and innovation program under grant agreement No 957189 (BIGMAP). The authors also thank CIDETEC for the electrode manufacturing. Finally, Joël Lachambre is also thanked for the Fiji plugin used to estimate geometric tortuosity.

\section*{Declarations}

The authors declare no Conflict of interest/Competing interests.



\end{document}